\documentclass[sigconf,screen]{acmart}

\startPage{1}
\setcopyright{none}

\usepackage{amsmath,amsfonts,bbm}
\usepackage{graphicx}
\usepackage{textcomp}
\usepackage{xcolor}
\usepackage{fancyhdr}
\usepackage{enumitem}
\usepackage{hyperref}
\usepackage{rotating}
\usepackage{booktabs}
\usepackage{balance}
\usepackage{algorithm}
\usepackage[noend]{algpseudocode}
\usepackage{multirow}
\usepackage{soul}
\usepackage{blkarray}
\usepackage{lipsum,wrapfig}
\usepackage{caption}
\usepackage{subcaption}
\usepackage{tikz}
\usepackage{cleveref}

\crefformat{section}{\S#2#1#3} 
\crefformat{subsection}{\S#2#1#3}
\crefformat{subsubsection}{\S#2#1#3}

\algnewcommand\algorithmicforeach{\textbf{for each}}
\algdef{S}[FOR]{ForEach}[1]{\algorithmicforeach\ #1\ \algorithmicdo}

\setlist{noitemsep, leftmargin=*, topsep=1pt, partopsep=0pt}

\newcommand{\SYSTEM}{SCOPE}
\newcommand{\Static}{Static}
\newcommand{\Oracle}{Oracle}
\newcommand{\Offline}{Offline}
\newcommand{\RAPL}{RAPL}
\newcommand{\BO}{BO}
\newcommand{\Stage}{StageOPT}
\newcommand{\NO}{SCOPE-NO}
\newcommand{\MLP}{SCOPE-MLP}
\newcommand{\LIN}{SCOPE-LINEAR}
\newcommand{\RF}{SCOPE-RF}

\newcommand{\x}{{\mathbf{x}}}

\DeclareMathOperator*{\argmax}{arg\,max}
\DeclareMathOperator*{\argmin}{arg\,min}

\newcommand*\circled[1]{\tikz[baseline=(char.base)]{
		\node[shape=circle,draw,inner sep=0.7pt] (char) {#1};}}

\definecolor{tab-blue}{RGB}{31, 119, 180}
\definecolor{tab-orange}{RGB}{255, 127, 14}
\definecolor{tab-red}{RGB}{214, 39, 40}

\begin{document}
	
\title{\SYSTEM{}: Safe Exploration for Dynamic Computer Systems Optimization}

\author{Hyunji Kim}
\affiliation{
	\institution{MIT EECS}
	\city{Cambridge, MA}
	\country{USA}}
\email{hyunjik@mit.edu}

\author{Ahsan Pervaiz}
\affiliation{
	\institution{University of Chicago}
	\city{Chicago, IL}
	\country{USA}}
\email{ahsanp@uchicago.edu}

\author{Henry Hoffmann}
\affiliation{
	\institution{University of Chicago}
	\city{Chicago, IL}
	\country{USA}}
\email{hankhoffmann@cs.uchicago.edu} 

\author{Michael Carbin}
\affiliation{
	\institution{MIT CSAIL}
	\city{Cambridge, MA}
	\country{USA}}
\email{mcarbin@csail.mit.edu}

\author{Yi Ding}
\affiliation{
	\institution{MIT CSAIL}
	\city{Cambridge, MA}
	\country{USA}}
\email{ding1@csail.mit.edu}

\begin{abstract}	
 
  Modern computer systems need to execute under strict safety
  constraints (e.g. a power limit), but doing so often conflicts with
  their ability to deliver high performance (i.e. minimal latency).
  Prior work uses machine learning to automatically tune hardware
  resources such that the system execution meets safety constraints
  optimally.  Such solutions monitor past system executions to learn
  the system's behavior under different hardware resource allocations
  before dynamically tuning resources to optimize the application
  execution.  However, system behavior can change significantly
  between different applications and even different inputs of the same
  applications. Hence, the models learned using data collected a
  priori are often suboptimal and violate safety constraints when used
  with new applications and/or inputs.
  
  To address this limitation, we introduce the concept of an
  \emph{execution space}, which is the cross product of hardware
  resources, input features, and applications. Thus, a
  \emph{configuration} is defined as a tuple of hardware resources,
  input features, and application. To dynamically and safely allocate
  hardware resources from the execution space, we present
  \SYSTEM{}~\footnote{\textbf{S}afe \textbf{C}onfiguration
    \textbf{O}ptimization for \textbf{P}erformance and
    \textbf{E}fficiency}, a resource manager that leverages a novel
  safe exploration framework. \SYSTEM{} operates iteratively, with
  each iteration (i.e., reallocation) having three phases: monitoring,
  safe space construction, and objective optimization. To construct a
  safe set with high coverage (i.e., a high number of safe
  configurations in the predicted safe set), \SYSTEM{} introduces a
  locality preserving operator so that \SYSTEM{}'s exploration will
  rarely violate the safety constraint and have small magnitude
  violations if it does.  We evaluate \SYSTEM{}'s ability to deliver
  improved latency while minimizing power constraint violations by
  dynamically configuring hardware while running a variety of Apache
  Spark applications.  
  Compared to prior approaches that minimize power constraint violations, \SYSTEM{} consumes comparable power while improving latency by up to 9.5$\times$. Compared to prior approaches that minimize latency, \SYSTEM{} achieves similar latency but reduces power constraint violation rates by up to 45.88$\times$, achieving almost zero safety constraint violations across all applications.

\end{abstract}

\maketitle

\section{Introduction}

A modern computer system needs to meet conflicting goals---for
example, minimizing latency while meeting some safety constraint
(e.g., power limit)---in the face of dynamic changes in its
application and environment. To do so, hardware architects expose a
wide variety of resources for the system to
manage~\cite{deng2017memory,rahmani2018spectr,zhang2021sinan}, where
each type of hardware resource is controlled by a \emph{hardware
  parameter} and all possible allocations of hardware resources
constitute a \emph{resource space}.

When managing these resources, the system needs to meet its safety
constraints.  Rare and small magnitude violations can be tolerable if
they do not dramatically degrade system
performance~\cite{yuan2003energy,fan2016computational,CALOREE}.
However, frequent and large magnitude violations can cause serious
damage, and even crash the system~\cite{raghavendra2008no}. For
example, power capping systems deployed in hyperscale datacenters
(e.g., Amazon, Google, Microsoft) smooth out spikes from occasional
power overloading~\cite{sakalkar2020data,li2020thunderbolt}, but
cannot tolerate large-scale violations without causing significant
degradation in application performance~\cite{ahmad2010joint}.

To ensure that the system executes optimally while meeting the safety
constraint, existing resource managers use \emph{samples from the resource space}
(i.e., hardware parameters and their measured system behavior) of past
executions to model system behavior (e.g., power and performance) as a
function of hardware resource usage. The model is then used to
dynamically adjust resources usage such that safety is maintained and
application performance is
optimized~\cite{ipek2006efficiently,LEO,CALOREE,hoffmann2015jouleguard,ding2019generative}.
However, samples collected from the resource space may not be
generalizable across different applications and inputs. This means
that safe and high-performing hardware resource allocation for one execution
can be unsafe and low-performing in
another~\cite{ding2021generalizable, yu2018datasize}.  Specifically,
when a new application or input leads to significantly different
system behavior, models learned using past executions cannot provide
safety guarantees and optimal performance.

\textbf{Execution space.} To address this limitation , we introduce
\emph{execution space}, which is the cross product of hardware
resources, input features, and applications.
We define a \emph{configuration} as a tuple of hardware
resources, input features (e.g., data size), and application.
To ensure that the system executes optimally while meeting the safety
constraint, we need to \emph{explore} (i.e., evaluate a configuration 
that the system has not seen before) the execution space, rather than
the resource space. Exploring the execution space is the process of
evaluating a previously unseen configuration from the execution space
and it allows us to learn models of system behavior as a function of
both the hardware parameters, the current application, and input.

\textbf{Our solution: safe exploration.} To dynamically and safely
allocate hardware resources from the execution space, we present
\SYSTEM{}, a resource manager that leverages a novel safe exploration
framework.  \emph{Safe exploration} is a family of sequential
decision-making techniques that optimize an objective while minimizing
safety constraint
violations~\cite{sui2015safe,sui2018stagewise,turchetta2019safe}.
Unlike static configuration that uses the same hardware resources
throughout application execution, \SYSTEM{} dynamically reallocates
hardware resources to optimize system performance and minimize safety
constraint violations while responding to dynamic runtime changes.
\SYSTEM{} operates iteratively, with each iteration (i.e.,
reallocation) having three phases: \emph{monitoring, safe space
  construction, and objective optimization}.
\begin{itemize}
\item In the monitoring phase, \SYSTEM{} samples the execution space;
  i.e., it measures the system behavior for the current application,
  input, and hardware resource allocation. Specifically, it records
  both the objective and safety data, and checks whether the safety
  constraint has been violated or not.
\item Safe space construction is the process of predicting a
  \emph{safe set} (i.e., a set of configurations that will not violate
  the safety constraint) with high coverage (i.e., a high number of
  safe configurations in the predicted safe set).  Furthermore, if an
  unsafe configuration is included in the safe set, its violation
  magnitude should be small.  To construct a safe set with these
  properties, \SYSTEM{} introduces a locality preserving operator
  based on the \emph{locality preserving
    criterion}~\cite{belkin2003laplacian} (i.e., if two configurations
  are close in distance, their system behavior is likely close as
  well). This operator constrains \SYSTEM{} to explore only
  configurations within a certain distance of the most recently executed
  safe configuration.  Because any new configurations are close to
  known safe configurations, exploring these new configurations will
  rarely violate the safety constraint and have small magnitude
  violations if it does.
\item In the objective optimization phase, \SYSTEM{} reallocates
  hardware resources with the best predicted performance from the
  newly constructed safe set.
\end{itemize}

\textbf{Results and contributions.} We evaluate \SYSTEM{}'s ability to
minimize latency (the objective) while meeting a power (the safety)
constraint for a variety of Apache Spark applications~\cite{spark}.
For each input and application, \SYSTEM{} dynamically configures
hardware resources (e.g., CPU frequency, uncore frequency, number of
sockets, number of cores per socket, and whether hyperthreads are
enabled).  Compared to prior approaches that minimize power constraint
violations, we find:
\begin{itemize}	
\item Compared to Intel's \RAPL{}~\cite{david2010rapl}, \SYSTEM{}
  decreases the violation rate by 54.1$\times$ and violation magnitude
  by 1.04$\times$. These violation reductions occur because \SYSTEM{}
  can reach even lower power caps than \RAPL{}.  Since \RAPL{} does
  not optimize latency, \SYSTEM{} is able to decrease latency by
  9.5$\times$.
\item Compared to an existing state-of-the-art safe exploration
  approach from domains outside of computer systems, \SYSTEM{}
  decreases latency by 1.11$\times$ across all evaluated applications
  while decreasing the violation rate and magnitude by 11.93$\times$
  and 1.40$\times$ respectively.  \SYSTEM{} achieves these results
  because it accounts for the unique features of computer systems by
  continually monitoring the system; prior work assumes that samples
  taken early in execution accurately capture behavior over the system
  lifetime.
\item With its locality preserving operator, \SYSTEM{}'s safe set has
  1.96$\times$ higher coverage and 1.35$\times$ lower violation
  magnitude than \NO{}, a version of \SYSTEM{} that does not use the
  operator (\cref{sec:res-op}), which means that even in the rare case
  where an unsafe configuration is selected from \SYSTEM{}'s safe set,
  it will likely have lower violation magnitude than that of \NO{}'s.
\end{itemize}
We summarize the contributions as follows:
\begin{itemize}
\item Expanding the exploration space from resource space to execution space, which captures
 that system behavior is function of hardware resources,
  application, and input.
\item Presenting \SYSTEM{}, a resource manager that leverages the safe
  exploration framework to optimize the objective while minimizing the
  safety constraint violations.
\item Introducing the locality preserving operator to construct the
  safe set with high coverage.
\end{itemize}
Safe exploration is an important, emerging frontier of machine
learning with a wealth of applications in safety-critical systems. To
the best of our knowledge, \SYSTEM{} is the first demonstration of
safe exploration in execution space for computer systems optimization.
\SYSTEM{} outperforms existing safe exploration techniques from other
domains by developing a locality preserving operator and requiring fewer assumptions about
application behavior than other safe exploration techniques. Our work
offers a foundation on which the computer systems community can build
new optimization tools that aid in exploration (to improve system
performance) while preserving safety.

\section{Related Work}\label{sec:related} 

This paper's key insight is a methodology for using machine learning
to perform optimal resource management while meeting safety
constraints in execution space; i.e., without accounting for application 
and input. This section focuses on related
work in machine learning for computer systems optimization
(\cref{sec:related-ml-uncon}, \cref{sec:related-ml-con}).  Rather than
learning the relationships between all three components (i.e.,
hardware resources, input, and application) from the execution space,
prior work learns from the resource space by assuming that samples of
the past executions from the resource space can accurately predict the
system behaviors of future executions from potentially different
applications or inputs.  We also explain the difference between
existing safe exploration techniques (from domains other than computer
systems) and \SYSTEM{} (\cref{sec:related-safe}).

\subsection{Machine Learning for Unconstrained Optimization}\label{sec:related-ml-uncon}

Machine learning techniques have been increasingly applied to solve
computer systems optimization problems by modeling complex, nonlinear
relationships between system resource usage and quantifiable
behavior~\cite{penney2019survey,ZhangH2019learned}. Much prior work
focuses on unconstrained optimization problems with no safety
constraint---i.e., optimizing a single objective such as
latency~\cite{belay2014ix}, throughput~\cite{li2020thunderbolt},
power~\cite{lee2006accurate}, and energy
consumption~\cite{yuan2003energy}. These works share a common
methodology of building machine learning models using training data
collected by sampling the system resource space: allocating different
system resources and measuring behaviors. Specifically, there are two
types of system sampling, random and intelligent, and each leads to
different types of machine learning approaches. Random sampling
typically needs a large amount of samples to build a highly accurate
model, but it is free of biases that might arise from intelligent
sampling~\cite{yi2003statistically,lee2006accurate,bitirgen2008coordinated,ansel2011language,chen2011modeling,cochran2011pack,oliner2013carat,deng2017memory,ansel2012siblingrivalry,ponomarev2001reducing,sridharan2013holistic,garza2019bit,bhatia2019perceptron,shi2019applying,shi2021hierarchical}.
For example, \citet{lee2008cpr} build a regression model on simulated
data to predict multiprocessor performance.  Paragon~\cite{paragon}
and Quasar~\cite{quasar} apply collaborative filtering to predict QoS
performance of workloads in datacenters.  However, the large sampling
effort required for random sampling is often prohibitive due to is
high computational cost and inefficiency~\cite{yu2018datasize}.

Unlike random sampling, intelligent sampling significantly reduces the
sampling effort required to achieve good systems
outcomes~\cite{ipek2006efficiently,Ipek2008self,petrica2013flicker,venkataraman2016ernest,canino2018stochastic}.
A representative family of intelligent sampling techniques is Bayesian
optimization, which iteratively samples data points that are predicted
to contribute the most information to the learning
model~\cite{frazier2018tutorial}.
CherryPick~\cite{alipourfar2017cherrypick} and
CLITE~\cite{patel2020clite} use Bayesian optimization to schedule
workloads in datacenters. GIL~\cite{ding2021generalizable} and
Bliss~\cite{roy2021bliss} use Bayesian optimization to optimize system
performance. HyperMapper~\cite{nardi2019practical} and
BOCA~\cite{chen2021efficient} use Bayesian optimization to tune
compilers. These works reduce the number of samples required to
perform optimization, but do not consider any safety constraints.
This motivates the need for a learning-based resource manager that
both (1) works with reduced number of samples and (2) respects safety
constraints.

\subsection{Machine Learning for Safe Optimization}\label{sec:related-ml-con}

Safe optimization problems in computer systems find the optimal
point within a tradeoff space (e.g., performance versus power)---i.e.,
optimizing a performance objective under some safety
constraint~\cite{li2006dynamic,dubach2010predictive,hoffmann2015jouleguard,deng2017memory,ding2019generative}.
For example, \citet{li2006dynamic} collect samples to optimize power
under a performance constraint.  \citet{dubach2010predictive} collect
samples to build a dynamic control system that optimizes energy and
performance efficiency. LEO~\cite{LEO} and CALOREE~\cite{CALOREE}
develop hierarchical Bayesian models to meet latency constraints and
minimize energy.

These works provide safety under the assumption that the samples they
collect from the resource space of past executions can accurately
capture the system behaviors that will be seen during future
executions.  However, this assumption can be violated when a new
application or new input causes significantly different system
behaviors than those in the collected samples. The emergence of such
unsampled behavior would render the whole system unsafe. As such,
there is a need for approaches that can explore unseen configurations
in the execution space, rather than the resource space.

\subsection{Safe Exploration in Other Problem Domains}\label{sec:related-safe}

Recent years have witnessed safe exploration applied to
safety-critical domains such as autonomous
driving~\cite{razin2020hitting}, healthcare~\cite{sui2018stagewise},
and robotics~\cite{brunke2021safe}. Safe exploration is a family of
sequential decision-making techniques that optimize an objective while
minimizing safety constraint
violations~\cite{garcia2012safe,sui2015safe,sui2018stagewise,turchetta2019safe,mao2019towards,turchetta2020safe,wachi2020safe,xu2021safely}.
To satisfy some safety property with a high probability, these
techniques either require extra supervision or knowledge accumulated
before exploration.  They achieve probabilistic (i.e.,
not deterministic) safety guarantees based on the assumption that the
changes of safety measurements are continuous and bounded (i.e.,
Lipschitz continuity). In computer systems, however, this assumption
does not hold. A typical example is power, where the power usage can
change dramatically every second~\cite{zhang2015quantitative}.
Different from prior work, our solution, \SYSTEM{}, is a new safe
exploration framework that introduces a locality preserving operator
to eliminate the need for such assumptions.  Evaluation results show
that \SYSTEM{} outperforms the existing state-of-the-art safe
exploration technique (e.g., \Stage{}~\cite{sui2018stagewise}) in both
performance and safety by tailoring our approach to the unique
properties of computer systems.

\section{Motivational Examples}\label{sec:motivation}

We use two examples to demonstrate that to explore new configurations safely, optimization must operate in execution space rather than resource space because safe
samples from the resource space of past executions do not generalize
to new applications and new inputs. To demonstrate a lack of
generalization across applications, we run two different applications
and show that the safe samples from the resource space for one
application are no longer safe for another. To demonstrate a lack of
generalization across inputs, we run one application with two
different inputs and show that the safe samples from the resource
space for one are not safe for the other.

We consider two Apache Spark applications from
HiBench~\cite{huang2010hibench} and run them on the Chameleon
configurable cloud computing platform~\cite{keahey2020lessons}
(details in \cref{sec:hw-systems}). We collect samples by profiling
the application at all possible assignments of hardware parameters
(see Table~\ref{tbl:hw-config}) and recording their latency and power
data.  Our goal is to minimize latency while meeting a power
constraint.

\textbf{Safe samples from the resource space do not generalize across
  applications.} We show that safe samples from the resource space do
not generalize across the applications \texttt{als} and
\texttt{nweight}.  We construct a safe set for \texttt{als} by
randomly selecting a list of hardware allocations that do not violate
the safety constraint based on all samples collected for \texttt{als}.
Then, we dynamically configure \texttt{als} (at 20s intervals) using
this safe set.  We then use the same safe set to dynamically configure
\texttt{nweight} (again, at 20s intervals).

\begin{figure}[t]
	\centering
	\includegraphics[width=\linewidth]{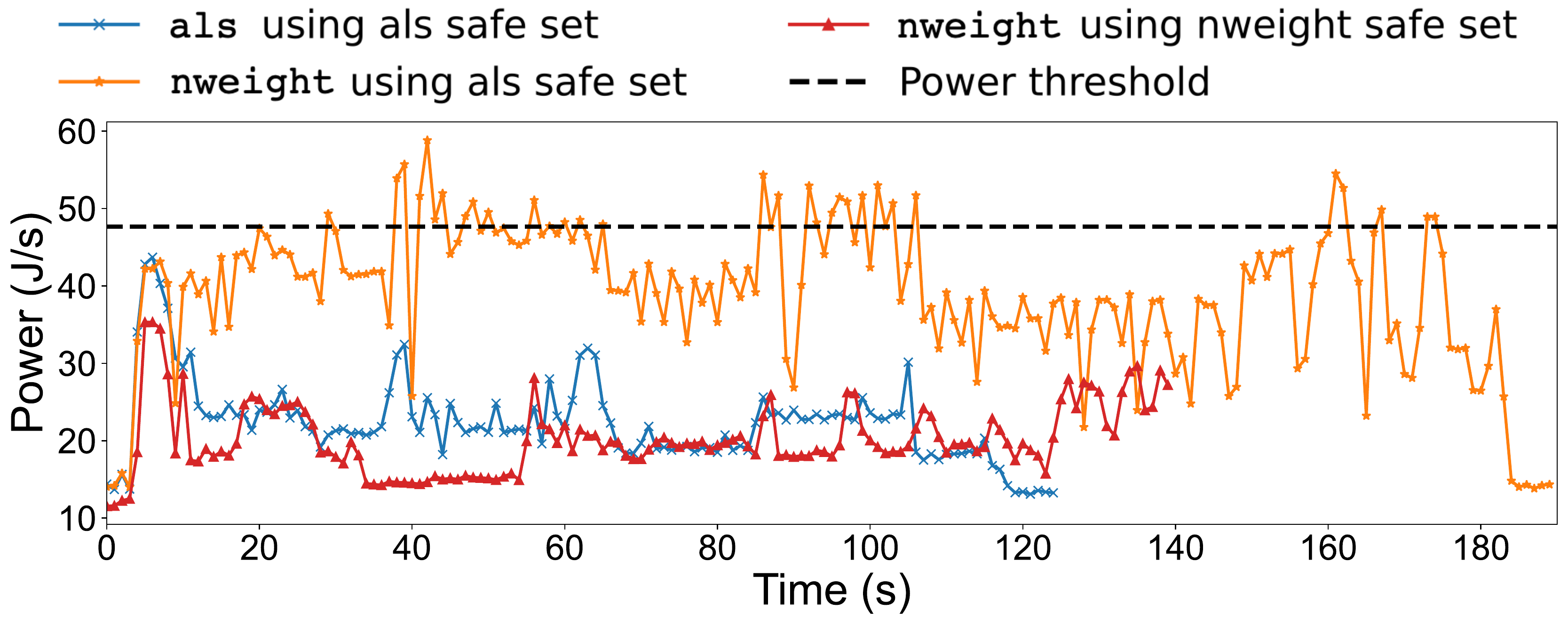} 
	\caption{Results of power and latency for executing \texttt{als} and \texttt{nweight}.}\label{fig:motivation_app}
\end{figure}

Figure~\ref{fig:motivation_app} shows the resulting power as a
function of execution time, where the x-axis is the execution time,
and the y-axis is the power.  The \textcolor{tab-blue}{blue} line
represents \texttt{als} using its own safe set, the
\textcolor{tab-orange}{orange} line represents running
\texttt{nweight} using \texttt{als}'s safe set, the
\textcolor{tab-red}{red} line represents running \texttt{nweight}
using safe hardware allocations constructed especially for
\texttt{nweight} (again by profiling all possible assignments of
hardware parameters). The black dotted horizontal line is the power
limit which represents the safety constraint in this example. We
observe that \texttt{als} finishes safely in 125s with 0 power
violations, while \texttt{nweight} using \texttt{als}'s safe hardware
allocations needs 190s despite 32 power violations.

The high number of violations and the significant difference in
latency, show that the safe samples from the resource space do not
generalize across different applications. In fact,
\texttt{nweight}---with a properly constructed safe set---finishes in
142s with 0 violations, indicating that it is possible to improve
\texttt{nweight}'s latency and safety, however, exploration operating only within the resource space fails to do so.

\textbf{Safe samples from the resource space do not generalize across
  different inputs for the same application.}  We show that safe
samples from the resource space do not generalize across different
inputs for the same application using \texttt{als}. We create two
different inputs with the same sizes (so the execution behavior
variations are due to properties of the data rather than data size):
\texttt{Input A} is the input we will sample to build a safe set and
\texttt{Input B} is the target input to be optimized.  We construct a
safe set for \texttt{Input A} by randomly selecting a list of hardware
allocations that do not violate constraints based on all samples we
collected for \texttt{Input A}.  Then, we dynamically configure for
\texttt{Input A} (at 20s intervals) using this safe set, and then
configure using the same safe set for \texttt{Input B} (again, at 20s
intervals).

\begin{figure}[t]
	\centering
	\includegraphics[width=\linewidth]{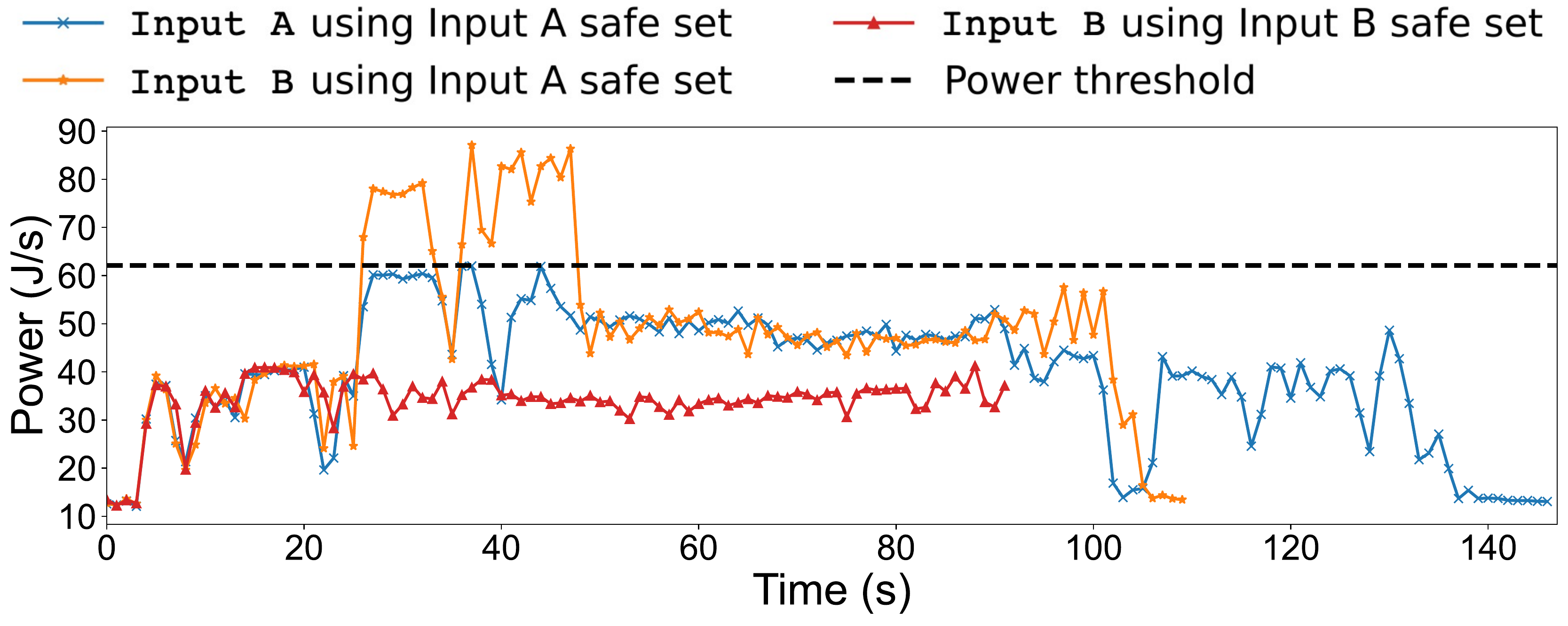}
	\caption{Results of power and latency for executing \texttt{als} with two different sets of inputs with same size.}\label{fig:motivation_input}
\end{figure}

Figure~\ref{fig:motivation_input} shows the the resulting power as a
function of execution time, where the x-axis is the execution time,
and y-axis is the power. The \textcolor{tab-blue}{blue} line
represents running \texttt{Input A} using its own safe set, the
\textcolor{tab-orange}{orange} line represents running \texttt{Input
  B} using \texttt{Input A}'s safe set, the \textcolor{tab-red}{red}
line represents running \texttt{Input B} with its own safe set found
by evaluating \texttt{Input B} in all possible assignments of hardware
parameters. The black dotted horizontal line is the power threshold,
or safety constraint in this example. We observe that \texttt{Input A}
finishes in 147s with 0 violation.  Running \texttt{Input B} with
\texttt{Input A}'s safe set, however, finishes in 110s with 20
violations, mainly between 26s and 46s. \texttt{Input B} is mostly
under the power threshold, but its dynamic behavior causes it to
violate safety constraint, while \texttt{Input A} does not.

The high number of violations when running \texttt{Input B} shows that
safe samples from the resource space for \texttt{Input A} are no
longer safe with different inputs, even when the inputs are the same
size.  In fact, \texttt{Input B} can finish in 91s with 0 violations
with an appropriate safe set, which indicates that the safe samples
from the resource space for \texttt{Input A} do not generalize to
\texttt{Input B}.

These examples demonstrate that safe samples from the resource space
of past executions fail to generalize to new applications and new
inputs. As an alternative to generalizing safe samples to new
applications or inputs, the next section describes \SYSTEM{}, our
solution to optimize objectives and minimize safety constraint
violations in the execution space, rather than resource space.

\begin{figure*}[!htb]
	\centering
	\includegraphics[width=0.9\textwidth]{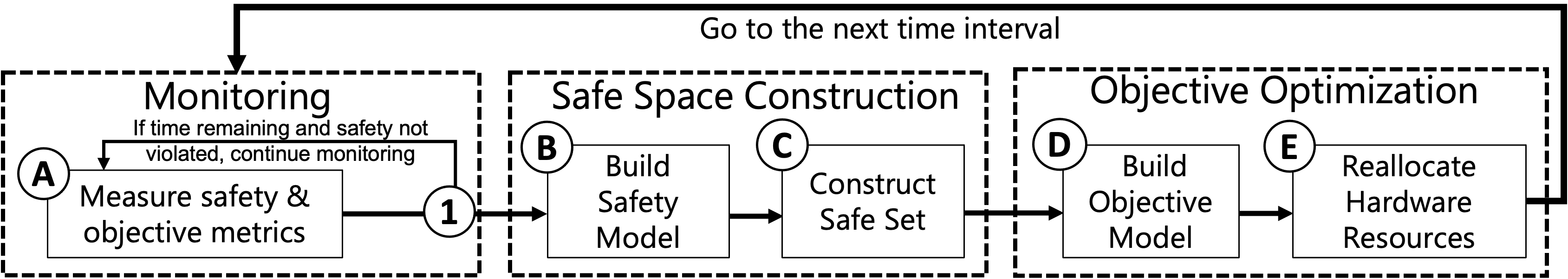}
	\caption{The workflow of \SYSTEM{} at each time interval while the application is executing. }
	\label{fig:workflow}
\end{figure*}

\section{\SYSTEM{} Design}\label{sec:framework}

\SYSTEM{} is a resource manager that dynamically and safely explores
in the execution space. In other words, \SYSTEM{} makes no assumptions
about how past executions, with different applications or inputs,
affect the current system behavior.  \SYSTEM{} iteratively reallocates
hardware resources, with each iteration having three phases:
monitoring, safe space construction, and objective optimization.
Figure~\ref{fig:workflow} illustrates \SYSTEM{}'s workflow. In the
first phase of each time interval, \SYSTEM{} continually measures
safety and objective metrics for the current application, input, and
hardware resource allocation (\circled{A}), and checks whether the
safety constraint has been violated or not (\circled{1}); if a safety
measurement violates the safety constraint, \SYSTEM{} moves to the
next phase immediately, otherwise it waits for a fixed time interval
to expire.  Then, \SYSTEM{} goes to the phase of safe space
construction to predict a safe set with high coverage. Within this
phase, \SYSTEM{} first builds a safety model using the measured
configurations and safety data (\circled{B}), and then constructs a
safe set based on the safety model (\circled{C}). Then, \SYSTEM{}'s
objective optimization phase reconfigures the system. Within this
phase, \SYSTEM{} first builds an objective model using the measured
configurations and objective data (\circled{D}), and then picks a
predicted high-performing configuration from the safe set and puts the
system into that configuration.

The remainder of this section first sets up the core concepts, and
then describes \SYSTEM{} in detail.

\subsection{Background and Definitions}

\SYSTEM{}'s input includes an application, objective metric to
optimize, safety constraint, a starting safe configuration, number of
measurements for each time interval, and a list of configurations over
which to optimize. A starting safe configuration is needed to prevent
\SYSTEM{} from violating the safety constraint during the first
iteration; users can conservatively choose this configuration. 
For example, if the safety metric is power, users could start
in a configuration with minimal hardware resources.  If latency is the
safety metric, then users could start with a configuration that makes
all resources available. \SYSTEM{}'s exploration will safely move the
system out of this conservative configuration to one that improves the
objective metric.

\begin{description}
\item[Application.] A program that runs on a computer system using hardware resources. 
\item[Configuration.] A configuration $\x_i\in D$ is a $d$-dimensional
  vector that includes $d$ parameters: $\x_i=[x_{i1}, x_{i2}, \ldots, x_{id}]$,
	where $i$ is the $i$-th configuration, and $x_{ij}$
        is the value for $j$-th parameter,
        $j\in[d]$.
      \item[Safety constraint.] The safety constraint is a threshold
        for a particular metric that the system does not want to
        violate during application execution. Some common metrics that can be
        used for safety constraints include
        energy~\cite{temam2012defect,hoffmann2015jouleguard},
        power~\cite{reagen2016minerva}, and
        latency~\cite{LEO,CALOREE,ding2019generative}. In this paper,
        we use $P$ to denote the safety constraint and $y_i$ the
        safety measurement for the $i$-th configuration.
      \item[Safe configuration.] The safe configuration is a
        configuration that does not violate the safety constraint,
        i.e., $y_i < P$. Similarly, an unsafe configuration is a
        configuration that violates the safety constraint, i.e.,
        $y_i\geq P$.
      \item[Optimization objective.]  The optimization objective is
        the metric that \SYSTEM{} minimizes or maximizes. Some
        common optimization objectives include
        energy~\cite{LEO,CALOREE,ding2019generative},
        power~\cite{reagen2016minerva}, and
        latency~\cite{temam2012defect,hoffmann2015jouleguard}. In this
        paper, we use $z_i$ to denote the optimization objective
        measurement for the $i$-th configuration.
      \item[Time interval.] The time interval is a short period of
        time during application execution. In this paper, we use $T_i$
        to denote $i$-th time interval when the application runs
        configuration $\x_i$. During each time interval, \SYSTEM{}
        will execute all three of its phases: monitoring, safe set
      construction, and objective optimization.
    \item[Measurement interval.] The measurement interval is the amount of time over which
      \SYSTEM{} gets a pair of safety and objective measurements.
\end{description}

\subsection{Monitoring}

At $i$-th time interval when the application runs at configuration
$\x_i$, \SYSTEM{} continuously gets safety $y_i$ and objective $z_i$ at each measurement interval for a maximum of $N$ times, where $N$ is a input parameter of \SYSTEM{} (\circled{A}). To react to violations as quickly as possible, if a safety measurement violates the safety constraint, \SYSTEM{} stops monitoring and moves to the next phase immediately instead of finishing $N$ measurements (\circled{1}). The measurement data collected over the time intervals are the training data for the next phases of \SYSTEM{}; i.e., until the $i$-th time interval, $X_{\rm train}= (\x_j)_{j=0}^i$, $Y_{\rm
	train}= (y_j)_{j=0}^i$, $Z_{\rm train}= (z_j)_{j=0}^i$.

\subsection{Safe Space Construction}

After getting the training data from measurement, \SYSTEM{} goes to
safe space construction, which is the process of predicting a safe 
set with high coverage---i.e., a
high number of safe configurations in the predicted safe set---and if
an unsafe configuration is included, its violation magnitude should be
small. To construct a safe set with these properties, \SYSTEM{}
introduces a \emph{locality preserving operator} based on the
\emph{locality preserving criterion}~\cite{belkin2003laplacian} (i.e.,
if two configurations are close in distance, their system behavior is
also likely to be close). The translation of this criterion to safe
space construction is as follows:
\begin{itemize}
\item If two configurations are close in distance, their corresponding
  safety and objective measurements are likely close too.
\item If configuration A is safe and configuration B is unsafe but
  close to configuration A in distance, the safety violation of
  configuration B is very likely to be small.
\end{itemize}
This criterion matches our empirical observation that a smaller
magnitude configuration change (e.g., changing 2 cores to 4 cores)
leads to smaller system behavior changes than a larger change (e.g.,
changing 2 cores to 12 cores). The locality preserving operator
constrains \SYSTEM{} to explore a subset of configurations in the
safe set that are within a neighborhood (i.e., close area within some
distance) of the most recently used safe configuration so that
\SYSTEM{}'s exploration will rarely violate the safety constraint and
have small magnitude violations if it does. Formally, $X_c$ is the candidate
set that is constructed with the locality preserving operator:
\begin{align}\label{eq:operator}
X_c = \{ \x \in D\setminus X_{\rm train} |~ \|\x - \x^s \| \leq \gamma\},
\end{align}
where $\x^s$ is the most recent safe configuration that \SYSTEM{} has been 
used, and $\gamma\geq 0$ is the operator parameter. The operator parameter
$\gamma$ controls the distance that \SYSTEM{} explores. If $\gamma=0$,
the only configuration that \SYSTEM{} can explore is the starting safe
configuration. If $\gamma>0$, \SYSTEM{} can explore configurations that are
within distance $\gamma$ from $\x^s$.  The higher $\gamma$ is, the larger 
configuration space \SYSTEM{} can explore, and thus the more likely the unsafe
configurations can be included in the safe set. When $\gamma$ is large
enough, it is equivalent to exploring without constraints.

Upon obtaining the candidate set $X_c$, \SYSTEM{} constructs the safe
set as follows. At $i$-th time interval $T_i$, \SYSTEM{} trains the
safety model $f_y$ using the configurations and safety
measurements collected so far $(X_{\rm train}, Y_{\rm train}) = (\x_j,
y_j)_{j=0}^i$ (\circled{B}). \SYSTEM{} uses this model to predict the
safety values of the configurations in the candidate set $X_c$, 
and include configurations that are predicted to meet the safety 
constraint in the safe set (\circled{C}):
\begin{align}
X_s = \{ \x \in X_c | f_y(\x) < P\},
\end{align}
where $X_s$ is the safe set, and $P$ is the safety constraint. The
design of safe space construction is compatible with any type of
learning models such as Gaussian process
regression~\cite{alipourfar2017cherrypick}, random
forest~\cite{nardi2019practical}, linear
model~\cite{ding2021generalizable}, and neural
networks~\cite{ipek2006efficiently}. \SYSTEM{} uses Gaussian process
regression since it performs the best in practice
(\cref{sec:res-surrogate}).

\subsection{Objective Optimization}

The objective optimization phase reallocates hardware resources with a
high-performing safe configuration. At $i$-th time interval, \SYSTEM{}
trains the objective model $f_z$ using the configurations and
objective measurements collected so far $(X_{\rm train}, Z_{\rm
  train}) = (\x_j, z_j)_{j=0}^i$ (\circled{D}).  \SYSTEM{} uses this
model to predict the objective values of the configurations from the
safe set. Empirically,
we find that the safe set could be empty when the operator parameter
$\gamma$ is too small, since there is not much neighborhood for
\SYSTEM{} to explore. To address the possible situation like this,
\SYSTEM{} does the following.
\begin{itemize}
\item If the safe set is empty, \SYSTEM{} picks the
  configuration in the candidate set $X_c$ that has the best predicted
  safety. In this way, \SYSTEM{} explores the configuration space 
  while avoiding safety violations.
\item If the safe set is not empty, \SYSTEM{} picks the
  configuration in the safe set $X_s = \{ \x \in X_r | f_y(\x) < P\}$
  that has the best predicted objective.
\end{itemize}
After \SYSTEM{} picks the new configuration and reallocates hardware sources based on this configuration (\circled{E}), it goes to the next time interval until the execution ends.

\begin{algorithm}[!htb]
	\small
	\caption{The SCOPE Resource Manager}\label{alg:scope}
	\begin{algorithmic}[1]
		\Require $\x_0$ \Comment{Starting safe configuration.} 
		\Require $P$ \Comment{Safety constraint threshold.}
		\Require $\gamma$ \Comment{Operator parameter.}
		\Require $N$ \Comment{Number of measurements.}
		\State $X_{\rm train} = \{\}$ \Comment{Set of sampled configurations.}
		\State $Y_{\rm train} = \{\}$ \Comment{Set for safety measurements.}
		\State $Z_{\rm train} = \{\}$  \Comment{Set for objective measurements.}
		\State $i = 0$
		\State $\x^s \gets \x_0$ \Comment{Assign current safe configuration.}
		\State Run application at $\x_0$. \label{line:start}
		\While{application running}
		\For {$t = 1 \text{,...} N$}
		\State Get safety $y_i$ and objective $z_i$. \label{line:measure}
		\If{$y_i > P$} \label{line:check-safety}
		\State Stop monitoring. \label{line:stop-measuring}
		\EndIf
		\EndFor
		\State Update training set $(X_{\rm train}, Y_{\rm train}, Z_{\rm train})$ with $(\x_i, y_i, z_i)$.  \label{line:add-data}
		\State Train safety model $f_y$ using $X_{\rm train}$ and $Y_{\rm train}$. \label{line:train-safety}
		\If{$y_i < P$} 
		\State $\x^s \gets \x_i$ \Comment{Update current safe configuration.} \label{line:current-safe}
		\EndIf 
		\State $X_c = \{ \x \in D\setminus X_{\rm train} |~ \|\x - \x^s \| \leq \gamma\}$ \Comment{Update candidate configuration set.}  \label{line:candidate-set}
		\State $X_s = \{ \x \in X_c | f_y(\x) < P\}$ \Comment{Construct safe set.} \label{line:safe-set}
		\State Train objective model $f_z$ using $X_{\rm train}$ and $Z_{\rm train}$. \label{line:train-objective}
		\If{$|X_s| == 0$} \label{line:empty-1}
		\State $\x_{i+1} \gets \argmin_{\x \in X_c} f_y(\x)$ \Comment{If safe set is empty, pick configuration with best predicted safety.}  \label{line:empty-2}
		\Else		
		\State $\x_{i+1} \gets \argmax_{\x \in X_s} f_z(\x)$ \Comment{If safe set is not empty, pick configuration with best predicted objective.} \label{line:not-empty}
		\EndIf 
		\State Reallocate hardware resources with $\x_{i+1}$. \label{line:reallocate}
		\State $i \gets i + 1$
		\EndWhile
	\end{algorithmic}
\end{algorithm}

\subsection{\SYSTEM{} Algorithm Summary}

\SYSTEM{} is a general system design compatible with any type of
learning models, safety constraints, and optimization objectives. To
demonstrate its effectiveness in solving systems problems, we use
\SYSTEM{} to minimize latency while constraining the power usage to be
below a fixed threshold, where the safety metric is power, and the
optimization objective is work done so far (e.g., total instruction
count). The input includes a starting safe configuration $\x_0$,  
safety constraint $P$, number of measurements $N$, and operator parameter $\gamma$. The
starting safe configuration can be set by the user since it is not
desirable that the application violates the safety constraint in the
beginning. The safety constraint depends on the optimization goal and
is set by the user. The operator parameter $\gamma$ is set by the user
to control the exploration space, where more insights can be found
in~\cref{sec:res-sensitivity}.

Algorithm~\ref{alg:scope} summarizes the procedure.
In~\cref{line:start}, \SYSTEM{} starts executing the application, and
records the current time $T_0$. While the application is executing,
\SYSTEM{} does the following steps iteratively.
For each measurement interval, \SYSTEM{} gets the safety and objective 
(\cref{line:measure}). If the safety measurement violates the safety constraint (\cref{line:check-safety}), \SYSTEM{} stops measuring and moves to the next phase (\cref{line:stop-measuring}). Otherwise, \SYSTEM{} continues measuring.
In~\cref{line:add-data}, \SYSTEM{} updates the training set by adding
the new measured data. In~\cref{line:train-safety}, \SYSTEM{} trains
the safety model using training configurations and safety
data. In~\cref{line:current-safe}, \SYSTEM{} updates the safe
configuration that has been most recently executed.
In~\cref{line:candidate-set}, \SYSTEM{} updates the candidate set by
applying the locality preserving operator.  In~\cref{line:safe-set},
\SYSTEM{} constructs the safe set from the candidate set based on the
predictions using the safety model. In~\cref{line:train-objective},
\SYSTEM{} trains the objective model using training configurations and
objective data. Then, \SYSTEM{} will conduct different
steps based on the cardinality of the safe set.
In~\cref{line:empty-1,line:empty-2}, if the safe set is empty,
\SYSTEM{} will pick the configuration with the best predicted safety
in the candidate set. Otherwise,
in~\cref{line:not-empty}, \SYSTEM{} will pick the configuration with
the best predicted objective in the safe set. 
In~\cref{line:reallocate}, \SYSTEM{} reallocates hardware resources based
on the newly pickled configuration. We
implement \SYSTEM{} in Python with libraries including
numpy~\cite{numpy}, pandas~\cite{pandas}, and
scikit-learn~\cite{scikit-learn}. The code is released
in~\url{https://anonymous.4open.science/r/scope-code-9999}.

\section{Experimental Setup} \label{sec:setup}

\subsection{Hardware System}\label{sec:hw-systems}

We experiment on the Chameleon configurable cloud computing
platforms~\cite{keahey2020lessons}, where each experiment runs on a
master node and four worker nodes. Each node is a dual-socket system
running Ubuntu 18.04 (GNU/Linux 5.4) with 2 Intel(R) Xeon(R) Gold 6126
processors, 192 GB of RAM, hyperthreads and TurboBoost. Each socket
has 12 cores/24 hyperthreads and a 20 MB last-level cache. We tune the
hardware parameters in Table~\ref{tbl:hw-config}. These parameters
have been shown to influence both latency and power tradeoffs and are
important to tune to optimally meet a power
cap~\cite{zhang2016maximizing}. In total, there are 1920 possible
allocations of hardware resources to be explored dynamically while minimizing
power cap violations.

\begin{table}[!ht]
	 \vspace{-0.1in}
	\begin{center}
		\caption{Hardware parameters.}
		\begin{tabular}{lccc}
			\toprule
			\textbf{Parameter} & \textbf{Range}  \\  \midrule
			CPU frequency (GHz)  & 1.0--3.7  \\ \hline
			Uncore frequency (GHz)  & 1.0--2.4  \\ \hline
			Hyperthreading & on, off \\ \hline
			Number of sockets & 1, 2 \\ \hline
			Number of cores per socket & 1--12 \\	\bottomrule
		\end{tabular}\label{tbl:hw-config}
	\end{center}
\vspace{-0.3in}
\end{table}

\subsection{Software System}\label{sec:sw-systems}

We use Apache Spark~\cite{spark} as our software system with the
default Spark configuration settings. We use 12 applications from
HiBench's benchmark suite~\cite{huang2010hibench}, which has been
widely applied to configuration optimization
evaluations~\cite{yu2018datasize,banerjee2021bayesperf,ding2021generalizable,wang2019tpshare}.
The applications cover various domains including microbenchmarks,
machine learning, websearch, and graph analytics (Table~\ref{tbl:applications}).

\begin{table}[!htb]
		\vspace{-0.1in}
	\begin{center}
		\caption{HiBench applications.} 
		\begin{tabular}{ll|ll} \toprule
			\textbf{Application}   & \textbf{Data size} & \textbf{Application} &  \textbf{Data size}  \\
			\midrule
			als         & 0.7 GB  & bayes       & 1 GB    \\
			gbt         & 0.1 GB    & kmeans      & 2.1 GB    \\
			linear      & 36 GB   & lr          & 2 GB     \\
			nweight     & 0.1 GB  & pagerank    & 0.2 GB   \\
			pca         & 0.1 GB    & rf          & 1.6 GB   \\
			terasort    & 3.2 GB  & wordcount   & 26 GB    \\
			\bottomrule   
		\end{tabular}
		\label{tbl:applications}
	\end{center} 
		\vspace{-0.2in}
\end{table}

\subsection{Points of Comparison}\label{sec:comparison}

We compare various approaches including static configuration, approaches
with extensive collected samples (\Offline{}), approaches without
collected samples (\RAPL{}, \BO{}, \Stage{}, \NO{}), and an approach
with perfect knowledge (\Oracle{}, which we create with brute force
search and is, of course, unrealizable in practice).

\begin{itemize}
\item \textbf{\Static{}}: run application at the starting safe
  configuration throughout the execution.
\item \textbf{\Offline{}}: before running the target application,
  randomly sample half of all possible hardware resource allocations using the target application and input, and build safety and objective models using Gaussian process regression. These models are not updated during application execution. We reconfigure using the predictions from these models at each time interval~\cite{LEO,CALOREE}.
\item \textbf{\RAPL{}}: Intel's Running Average Power Limit system
  that allows users to set a power limit and tunes processor behavior
  to respect that limit using dynamic voltage and frequency scaling technique~\cite{david2010rapl}; \RAPL{} only configures CPU and uncore frequency.
\item \textbf{\BO{}}: use Bayesian optimization to
  reconfigure at each time interval, with Bayesian Gaussian process 
  regression being the learning model and expected improvement being the acquisition
  function~\cite{alipourfar2017cherrypick,nardi2019practical,patel2020clite,chen2021efficient}.
\item \textbf{\Stage{}}: use the \Stage{} algorithm~\cite{sui2018stagewise}, 
  a representative of safe exploration approaches from other domains that assume continuous
  and bounded changes in safety measurements to achieve probabilistic
  safety guarantee.  \Stage{} separates safe set construction and
  objective optimization into two stages: in the first few iterations,
  it focuses on constructing the safe set only; and then it switches
  to optimizing the objective within the safe space without updating
  the safe set. The lack of update to the safe set can be very
  problematic in computer systems.  As shown in our motivational
  example, the discrete nature of computer systems means that
  applications that look safe might quickly change to unsafe with no
  prior warning (see Figure \ref{fig:motivation_input}).  In contrast,
  \SYSTEM{} continually updates its safe set.
\item \textbf{\NO{}}: use the introduced safe exploration framework without the
  locality preserving operator to reconfigure at each time interval.
\item \textbf{\SYSTEM{}}: use the introduced safe exploration framework with the
locality preserving operator to reconfigure at each time
  interval. Both \SYSTEM{} and \NO{} use Gaussian process regression
  as the learning model since it performs the best
  (\cref{sec:res-surrogate}).
\item \textbf{\Oracle{}}: profile all latency and power data for the 
entire configuration space, and pick the fastest configuration 
that meets the power constraint.
\end{itemize}

\subsection{Evaluation Methodology}

We start executing the application at the same safe configuration for
each approach in~\cref{sec:comparison}, and then reconfigure it
dynamically at each time interval. For all approaches that reconfigure
dynamically, we run the sweep over different time
intervals and pick the best time interval for each
(\cref{sec:res-interval}). During each time interval, we use the maximum power for safety and average of the work done (i.e. total instruction counts) for objective. 

We evaluate on a wide range of
power constraints that are set as $[40, 50, 60, 70, 80]$-th
percentiles of the power distributions that are achievable across all
configurations. This is a reasonable range in that a small constraint
value (e.g., $[10,20,30]$-th percentiles) leaves little room for
tuning and a large constraint value (e.g., $[90]$-th percentile) is
often too relaxed to constrain the power. 

To make a comprehensive comparison, we identify 5 starting configurations as
fast and 5 as slow. For fast configurations, the latency is below p35
of the latency distribution over all configurations that meet the
power constraint. For slow ones, the latency is above p65 of the
latency distribution that meet the power
constraint. The reported results are averaged over
different constraints and 10 different starting safe configurations.

For \SYSTEM{}, we choose the operator parameter $\gamma=1$
for all applications based on the best tradeoffs between the violation rates
and speedups from sensitivity analysis in \cref{sec:res-sensitivity}. 
We use 1 second as the measurement interval.

We note that \SYSTEM{} runs on the same hardware whose power it is
controlling.  Thus, \SYSTEM{} must account and compensate for its own
power overhead. All results include the power and latency overhead of
running \SYSTEM{}. 

\subsection{Evaluation Metrics}\label{sec:metric}

For latency evaluation, we measure the latency $l_{\rm confg}$
obtained by each approach, and compute its speedup compared to the
latency $l_{\rm static}$ of~\Static{}:
\begin{align}\label{eq:speedup}
\text{speedup} = \frac{l_{\rm static}}{l_{\rm confg}}.
\end{align}
For power evaluation, we record the power over 1 second interval. We
note the times that power exceeds the power threshold $n_{\rm
  violate}$, and then divide it by the total number of measurements
$n_{\rm total}$:
\begin{align}\label{eq:rate}
\text{violation~rate} = \frac{n_{\rm violate}}{n_{\rm total}} \times 100\%.
\end{align}
We also record the average of power usage that exceeds the power
threshold $p$ throughout application execution, and
then divide it by the power threshold $P$:
\begin{align}\label{eq:magnitude}
\text{violation~magnitude} = 
\begin{cases}
0,       & \quad \text{if } p \leq P\\
\frac{p}{P},  & \quad \text{if }  p > P
\end{cases}
\end{align}
For safe set, we define coverage of the safe set:
\begin{align}\label{eq:coverage}
\text{coverage} = \frac{m_{\rm safe}}{m_{\rm total}} \times 100\% ,
\end{align}
where $m_{\rm safe}$ is the number of true safe configurations in the
predicted safe set, and $m_{\rm total}$ is the number of all
configurations in the predicted safe set.  

\section{Experimental Evaluation}\label{sec:evaluation}

\begin{figure*}[!htb]
	\centering
	\includegraphics[width=\linewidth]{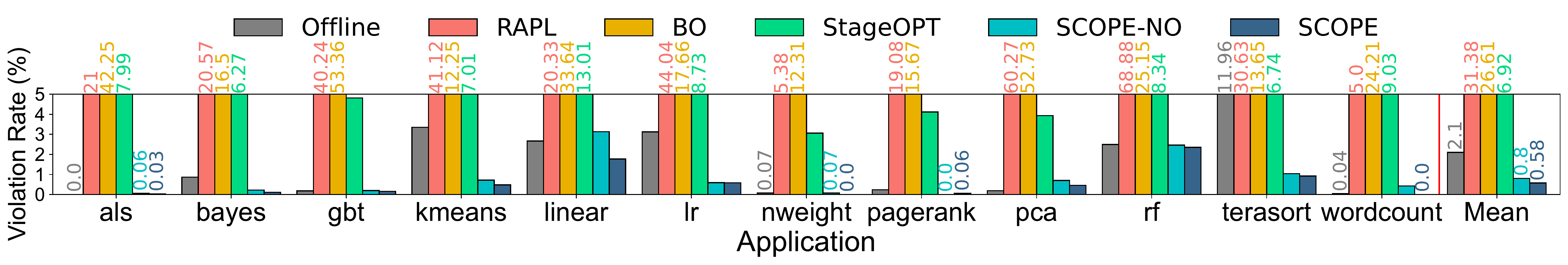}
	\caption{Violation rates averaged over all power constraints for each application. Lower is better.}\label{fig:violation-rate}
\end{figure*}

\begin{figure*}[!htb]
	\centering
	\includegraphics[width=\linewidth]{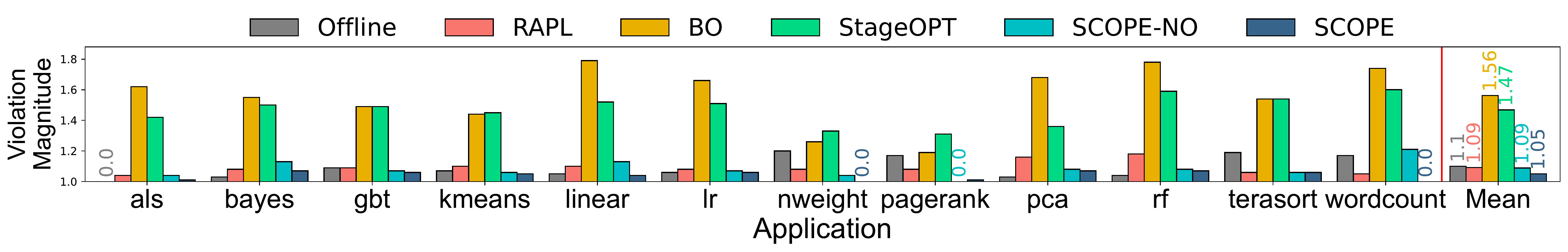} 
	\caption{Violation magnitudes averaged over all power constraints for each application. Lower is better.}\label{fig:violation-magnitude}
\end{figure*}

\begin{figure*}[!htb]
	\centering
	\includegraphics[width=\linewidth]{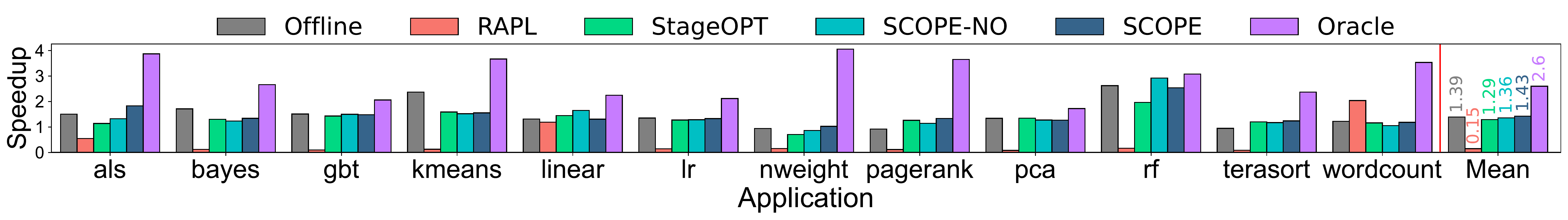} 
	\caption{Speedup averaged over all power constraints for each application. Higher is better.}\label{fig:latency}
\end{figure*}

\begin{figure*}[!htb]
	\centering
	\includegraphics[width=\linewidth]{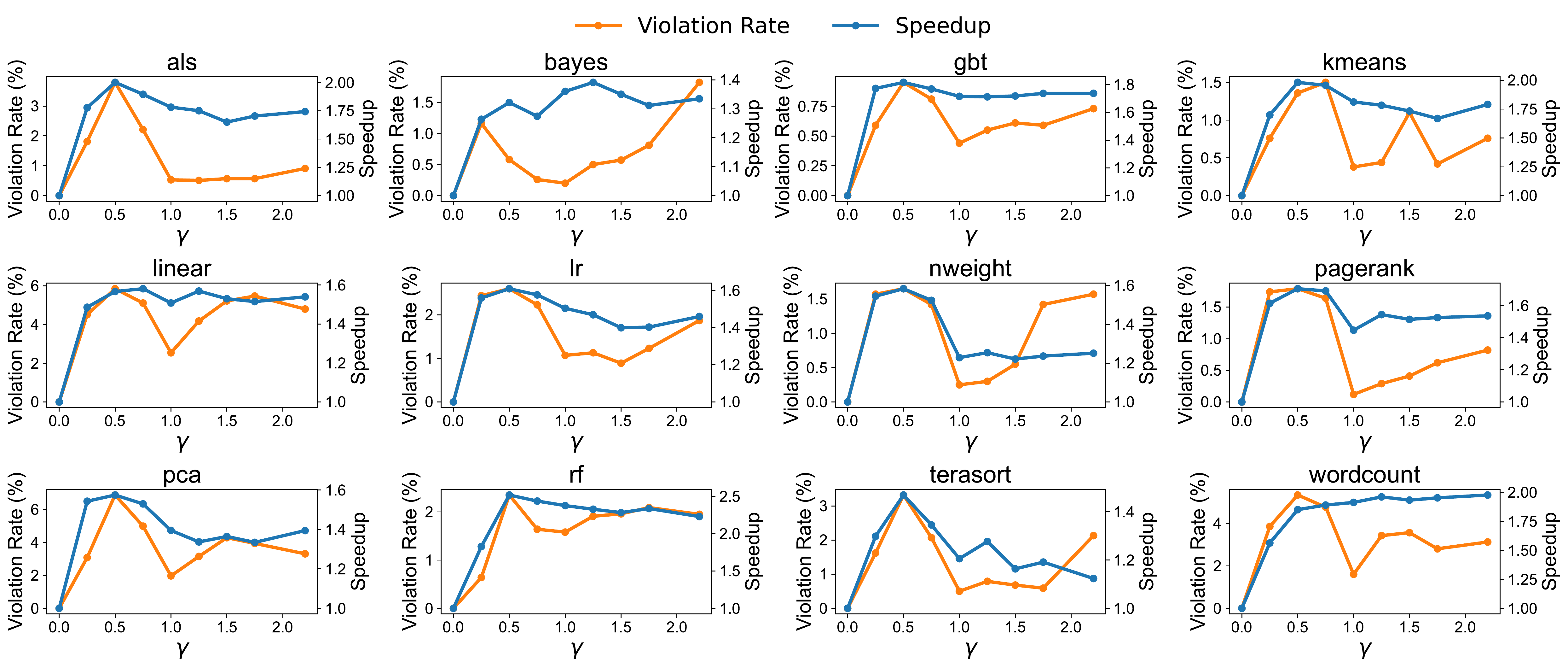} 
	\caption{Violation rate and speedup as a function of $\gamma$ averaged over all power constraints for each application. Lower is better for violation rate. Higher is better for speedup.}\label{fig:gamma}
\end{figure*}

We evaluate the following research questions (RQs):
\begin{itemize}
	\item \textbf{RQ1:} Does \SYSTEM{} reduce power violations? 
	\SYSTEM{} reduces violation rates by 3.62--54.1$\times$
		(Figure~\ref{fig:violation-rate}) and violation magnitudes by
		1.04--1.49$\times$ (Figure~\ref{fig:violation-magnitude}) compared
		to other baselines.
	\item \textbf{RQ2:} Does \SYSTEM{} improve application latency?
		\SYSTEM{} improves application latency by 1.07--9.5$\times$ compared to other baselines (Figure~\ref{fig:latency}).
    \item \textbf{RQ3:} Does locality preservation produce better safe sets?
    	With the locality preserving operator, \SYSTEM{}'s safe set has 1.96$\times$
    	higher coverage and 1.35$\times$ lower violation magnitude than
    	\NO{} that does not use the operator
    	(Table~\ref{tbl:safesetcounts}), which means that even in the rare
    	case where an unsafe configuration is selected from \SYSTEM{}'s safe
    	set, it will likely have lower violation magnitude than
    	that of \NO{}.
	\item \textbf{RQ4:} How sensitive is \SYSTEM{} to $\gamma$?
		The operator parameter $\gamma$ affects both speedup and
		violation rate, and all applications share a common trend of the tradeoffs between speedup and violation rate (Figure~\ref{fig:gamma}).
	\item \textbf{RQ5:} How sensitive is \SYSTEM{} to time intervals?
		\SYSTEM{} is the most robust to the time interval between reallocations compared to other baselines due to its ability to update models at all iterations and locality preserving operator (Figure~\ref{fig:time_interval_sensitivity}).
	\item \textbf{RQ6:} How do different types of models perform?
		\SYSTEM{} can be used with any types of learning models, and
		we have chosen Gaussian process regression for all of our evaluations due to
		its lowest violation rates and comparably low latency
		(Figure~\ref{fig:surrogates}).
	\item \textbf{RQ7:} What are the overheads?
		 \SYSTEM{} achieves low overhead of 0.05s per sample on
		average, which is negligible given the fact that we include the
		overheads in all experiments and \SYSTEM{} has the best latency
		improvement (Figure~\ref{fig:overhead}).
\end{itemize}

\begin{figure*}[!htb]
	\centering
	\includegraphics[width=\linewidth]{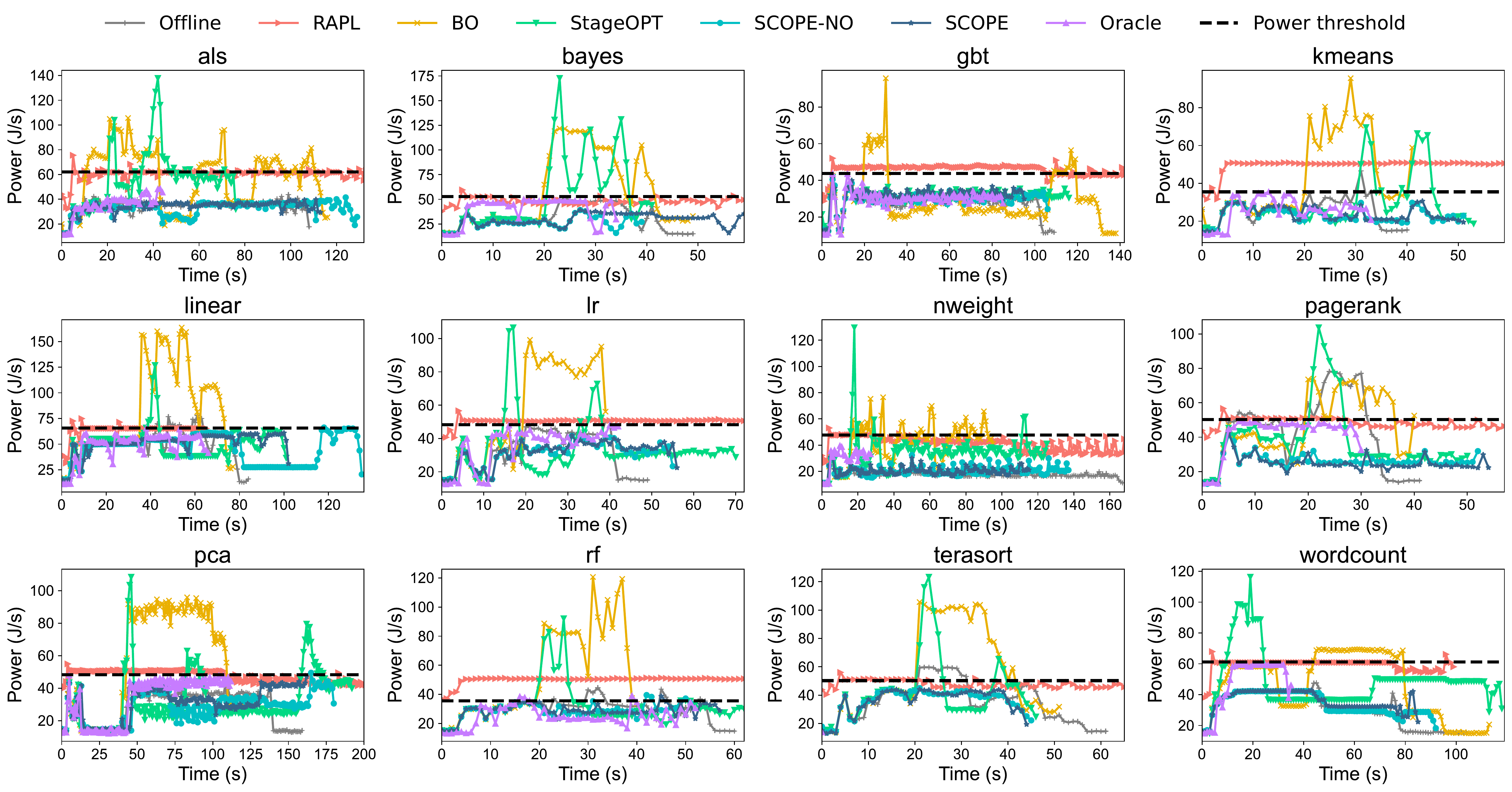} 
	\caption{Power results at different time points during execution for each application.}\label{fig:power_time_series}
\end{figure*}

\subsection{RQ1: Does \SYSTEM{} reduce power violations?}\label{sec:res-power}

We use violation rate and magnitude to evaluate how well \SYSTEM{}
reduces power violations. Figure~\ref{fig:violation-rate} and
\ref{fig:violation-magnitude} summarize the average violation rates
and magnitudes for different approaches, where the x-axis is the
application, the y-axis is the violation rate or magnitude and the
last column Mean is the arithmetic mean over all applications for Figure~\ref{fig:violation-rate} and the arithmetic mean over all applications with non-zero violation magnitude for Figure~\ref{fig:violation-magnitude}, where
we put numbers of each bar to quantify the average results. To
visualize better, we cap the violation rate at 5\% in
Figure~\ref{fig:violation-rate} and put numbers on the bars that are
capped. \Static{} and \Oracle{} were omitted since they do not violate
at all; \Static{} runs the same starting safe configuration throughout
the execution, and \Oracle{} runs the fastest configuration that is
safe based on the exhaustive search.
Figure~\ref{fig:power_time_series} shows power results at every
time point for different approaches, where x-axis is the running time, and y-axis is the
power. Overall, \SYSTEM{} has the lowest violation rate and magnitude,
making rare and small disruptions that disturb the system the least:
\begin{description}
\item[Violation rate.] \SYSTEM{} is 3.62$\times$ better than
  \Offline{}, 54.1$\times$ better than \RAPL{}, 45.88$\times$ better
  than \BO{}, and 11.93$\times$ better than \Stage{}.
\item[Violation magnitude.] \SYSTEM{} is 1.05$\times$ better than \Offline{}, 1.04$\times$ better than
  \RAPL{}, 1.49$\times$ better than \BO{}, 1.40$\times$ better than \Stage{}. 
\end{description}
Meanwhile, we observe the following:
\begin{itemize}
\item \Offline{} underperforms \SYSTEM{} in all but 2 applications.
  Despite utilizing a large amount of samples collected a priori, 
  the models \Offline{} builds a priori do not generalize to the environment changes that occur during execution, while \SYSTEM{} dynamically updates its
  models to reconfigure. As a result, \Offline{} has higher violation rate, which also demonstrates the
  difficulty of generalization of collected samples.
\item \RAPL{} has the highest violation rate despite the fact that 
it only optimizes for power. \RAPL{} achieves low 
  violation rate with higher power constraints (e.g. \texttt{nweight} and
  \texttt{terasort} in Figure~\ref{fig:power_time_series}), but the lower
  power constraints (i.e. [40,50]-th percentiles of the power
  distribution) are below the minimum power threshold that \RAPL{} can
  meet and cause \RAPL{} to violate constantly. This suggests that
  \SYSTEM{} performs much better than \RAPL{} with a wider range of constraints.
\item \BO{} has the second highest violation rate and largest
  violation magnitudes since it optimizes for
  latency only and does not consider power constraints.
\item \Stage{} underperforms \SYSTEM{} because \Stage{}
achieves probabilistic safety guarantees by assuming continuity and
boundedness of power measurements, which are not guaranteed to hold in computer systems. Thus \Stage{} makes inaccurate predictions, which leads it
to have higher violation rates and magnitudes.
\item \SYSTEM{} has 1.38$\times$ lower violation rate and 1.04$\times$
  lower violation magnitude than \NO{}. The results suggest that the
  introduced operator is beneficial for reducing violations (more
  details in \cref{sec:res-op}).
\end{itemize}

\begin{table*}[!htb]
\centering
\caption{Summarized results of the percentage of configurations (POC) selected in the total configuration space, coverage, and the average violation magnitudes (AVM) of unsafe configurations in the safe set for \NO{} and \SYSTEM{} when the application runs at the last time interval. Higher coverage is better. Lower AVM is better.}
\label{tbl:safesetcounts}
\begin{tabular}{@{}l|llllll@{}}
	\hline
	& \multicolumn{3}{c}{\NO} & \multicolumn{3}{c}{\SYSTEM{}} \\ \midrule
	& POC (\%) & Coverage (\%) & \multicolumn{1}{l|}{\begin{tabular}[c]{@{}l@{}}AVM 
	\end{tabular}} & POC (\%) & Coverage (\%) & \begin{tabular}[c]{@{}l@{}}AVM\end{tabular} \\ \hline
	als & 95.78 & 38.48 & \multicolumn{1}{l|}{1.47} & 19.13 & 71.08 & 1.16 \\
bayes & 95.71 & 39.77 & \multicolumn{1}{l|}{1.45} & 19.52 & 76.73 & 1.09 \\
gbt & 95.66 & 27.97 & \multicolumn{1}{l|}{1.61} & 20.81 & 55.88 & 1.16 \\
kmeans & 94.28 & 24.85 & \multicolumn{1}{l|}{1.66} & 18.31 & 56.04 & 1.19 \\
linear & 95.30 & 43.24 & \multicolumn{1}{l|}{1.42} & 20.08 & 76.32 & 1.12 \\
lr & 95.74 & 31.73 & \multicolumn{1}{l|}{1.52} & 18.60 & 65.47 & 1.10 \\
nweight & 95.30 & 31.63 & \multicolumn{1}{l|}{1.55} & 19.87 & 60.14 & 1.16 \\
pagerank & 95.83 & 39.92 & \multicolumn{1}{l|}{1.47} & 20.56 & 71.41 & 1.11 \\
pca & 93.06 & 30.19 & \multicolumn{1}{l|}{1.58} & 18.63 & 65.60 & 1.11 \\
rf & 92.21 & 19.13 & \multicolumn{1}{l|}{1.75} & 17.86 & 40.69 & 1.19 \\
terasort & 95.55 & 34.87 & \multicolumn{1}{l|}{1.51} & 19.04 & 74.32 & 1.12 \\
wordcount & 95.66 & 39.83 & \multicolumn{1}{l|}{1.47} & 18.74 & 73.61 & 1.19 \\ \midrule
Mean & 95.01 & 33.47 & \multicolumn{1}{l|}{1.54} & 19.26 & 65.61 & 1.14 \\ \bottomrule
\end{tabular}
\end{table*}

\subsection{RQ2: Does \SYSTEM{} improve application latency?}\label{sec:res-latency}

Figure~\ref{fig:latency} summarizes the speedups over \Static{} for
different approaches, where the x-axis is the application, the y-axis
is the speedup, and the last column Mean is the arithmetic mean over
all applications, where we put the number on each bar to quantify the
summarized results. We have omitted \BO{} as it optimizes for latency
only and thus has a very high violation rate and magnitude, which
makes it an unfair comparison. 

\Oracle{} has the highest latency
speedup since it has perfect prior knowledge to choose the fastest
safe configuration. \SYSTEM{} is second to \Oracle{}: 1.07$\times$
higher speedup than \Offline{}, 1.11$\times$ higher speedup than
\Stage{}, and 9.5$\times$ higher speedup than the slowest baseline,
\RAPL{}. In particular, we observe the following:
\begin{itemize}
\item All approaches outperform \Static{} except \RAPL{}, which
  demonstrates the effectiveness of dynamic reconfiguration during
  execution for optimizing latency. \RAPL{} is the exception
  because it manages power only and disregards latency.  In addition,
  because RAPL only configures core frequency, it is unable to take advantage of
  more complex tradeoffs that can reduce power without harming latency
  as much (for example, reducing core usage for applications with low
  parallel speedup)~\cite{zhang2016maximizing}.
\item \Offline{}, despite training on a large amount of samples collected prior to
  running the target application, has lower speedup than
  \SYSTEM{}. This is because \Offline{} uses
  fixed models trained over early collected data that fail to adapt to changes in execution.
\item \Stage{}, despite achieving good results in
  other problem domains, underperforms \SYSTEM{} for computer
  systems optimization. It is because \Stage{} focuses on constructing
  the safe set in the first stage and thus does not optimize latency
  until the second stage, while \SYSTEM{} constructs the safe set and
  optimizes latency throughout application execution.
\item \SYSTEM{} achieves 1.05$\times$ higher speedup than \NO{}. This
  suggests that the locality preserving operator not only reduces safety
  violations, but improves latency. Detailed analyses can be found in \cref{sec:res-op} and
  \cref{sec:res-sensitivity}.
\end{itemize}

\subsection{RQ3: Does locality preservation produce better safe sets?}\label{sec:res-op}

To better understand how the locality preserving
operator improves the safe set for \SYSTEM{} over \NO{} (the one
without the operator), we show the percentage of configurations (POC)
selected in the total configuration space, coverage, and the average
violation magnitudes (AVM) of unsafe configurations in the safe set
for \NO{} and \SYSTEM{} in Table~\ref{tbl:safesetcounts}, where the
numbers are averaged over different power thresholds and starting
configurations. \SYSTEM{}'s results are obtained when the operator
parameter $\gamma=1$ for all applications based on the best tradeoffs
between violation rate and speedup (\cref{sec:res-sensitivity}). We
observe the following:

\begin{itemize}
\item \SYSTEM{} has 4.9$\times$ POC smaller than \NO{}, which
  indicates that the operator significantly reduces the number of
  configurations included in the safe set.
\item Although \SYSTEM{}'s safe set has significantly fewer
  configurations, its coverage is 1.96$\times$ higher than that of
  \NO{}, which indicates that the operator greatly improves the
  accuracy of the safe set prediction. The higher accuracy that
  \SYSTEM{} has in predicting safe configurations leads \SYSTEM{} to
  achieve 1.38$\times$ lower violation rate than \NO{}
  (Figure~\ref{fig:violation-rate}).
\item \SYSTEM{} has 1.35$\times$ lower violation magnitude than \NO{}
  for all unsafe configurations in the safe set. This suggests that
  even when an unsafe configuration from \SYSTEM{} is chosen for
  reconfiguration, it will be likely to generate lower magnitude
  violation than that from \NO{}. This is reflected in
  Figure~\ref{fig:violation-magnitude} where \SYSTEM{} has the lowest
  overall violation magnitudes.
\end{itemize}

\subsection{RQ4: How sensitive is \SYSTEM{} to $\gamma$?}\label{sec:res-sensitivity}

The operator parameter $\gamma$ controls the size of neighborhood
space that \SYSTEM{} explores (Eq.~\ref{eq:operator}). To understand
the effects of $\gamma$ on \SYSTEM{}, we conduct sensitivity analysis of $\gamma$ on violation rate and speedup. For better visualization, we normalize each hardware parameter by mean normalization such that each parameter is within range $[-0.5, 0.5]$~\cite{alam2011comparative}. 
Figure~\ref{fig:gamma} shows violation rate and speedup as a function of
$\gamma$, where the x-axis is the different values of $\gamma$
and the y-axis is violation rate and speedup. Note that when $\gamma =
0$, this is equivalent to the \Static{} approach since \SYSTEM{} must
always choose the initial safe configuration. 
In particular, we observe the following:
\begin{itemize}
\item There is a tradeoff between violation rate and speedup; 
  lower violation rate is likely to have lower speedup
  and higher violation rate is likely to have higher speedup, where
  ideally, we want low violation rate and high speedup. For example,
  the average violation rate over all applications decreases from
  3.04\% (when $\gamma = 0.5$) to 0.93\% (when $\gamma = 1$), while
  average speedup of 1.75 (when $\gamma = 0.5$) also decreases to 1.60
  (when $\gamma = 1$). The tradeoff occurs because usually when a violation
  occurs, higher power is consumed, and more work
  is performed, allowing the application to finish earlier. 
\item The violation rate is generally higher when $ 0 < \gamma < 1$,
  compared to when $\gamma \geq 1$. This is because we normalize all
  hardware parameters to range from -0.5 to 0.5. Given binary
  parameters such as hyperthreading (on, off) and number of sockets
  (1, 2), \SYSTEM{} needs $\gamma$ to be at least 1 to consider
  updating those parameters. When $ 0 < \gamma < 1$, \SYSTEM{} is not
  enabled to change hyperthreading or number of sockets at all, which
  limits the search space for \SYSTEM{} to explore safely.
\end{itemize}

\begin{figure}[!htb]
	\centering
	\includegraphics[width=\linewidth]{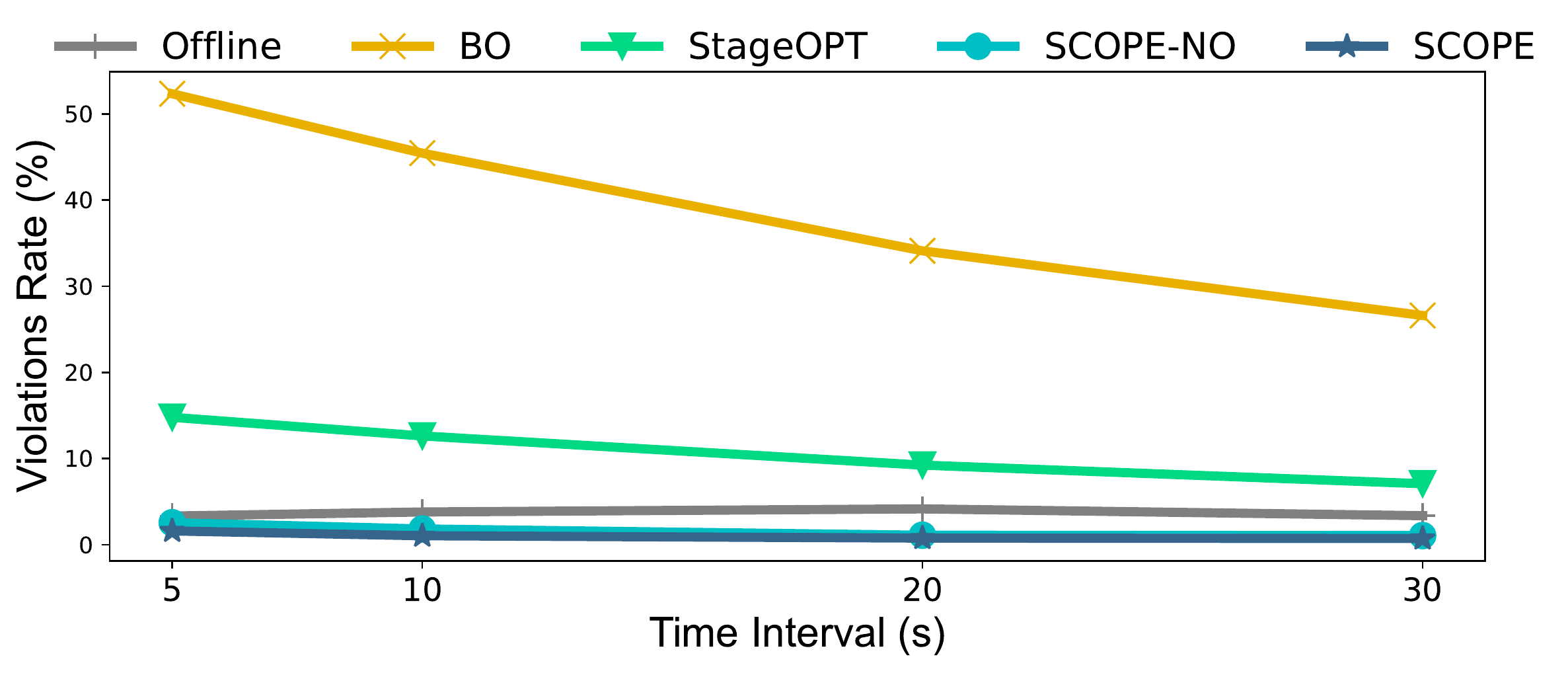} 
	\caption{Violation rate as a function of time interval averaged over all power constraints and applications for each approach.}\label{fig:time_interval_sensitivity}
	\vspace{-0.1in}
\end{figure}

\begin{figure*}[!htb]
	\centering
	\includegraphics[width=\linewidth]{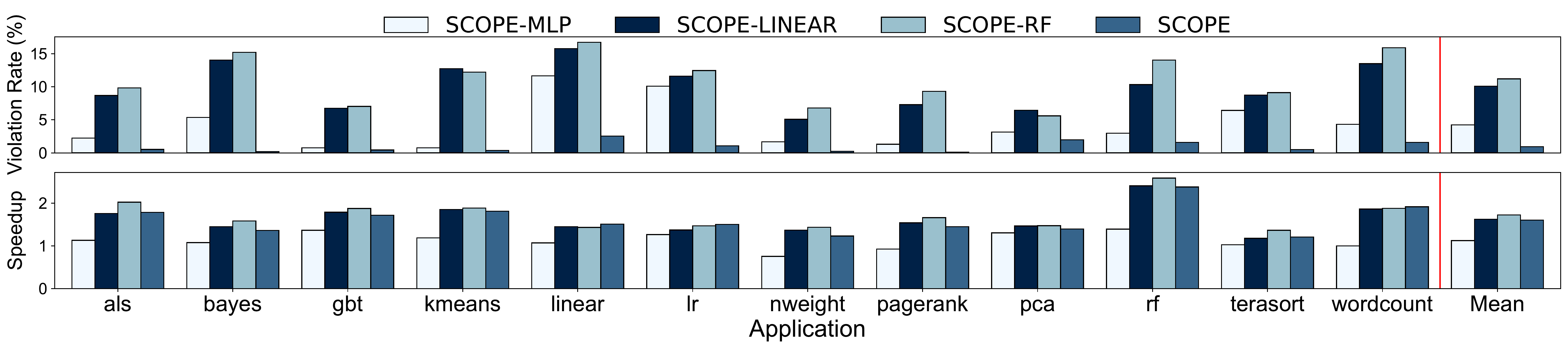} 
	\caption{Violation rate and speedup of \SYSTEM{} using different types of learning models, averaged over all power constraints for each application. Lower is better for violation rate. Higher is better for speedup.}\label{fig:surrogates}
\end{figure*}

\begin{figure*}[!htb]
	\centering
	\includegraphics[width=\linewidth]{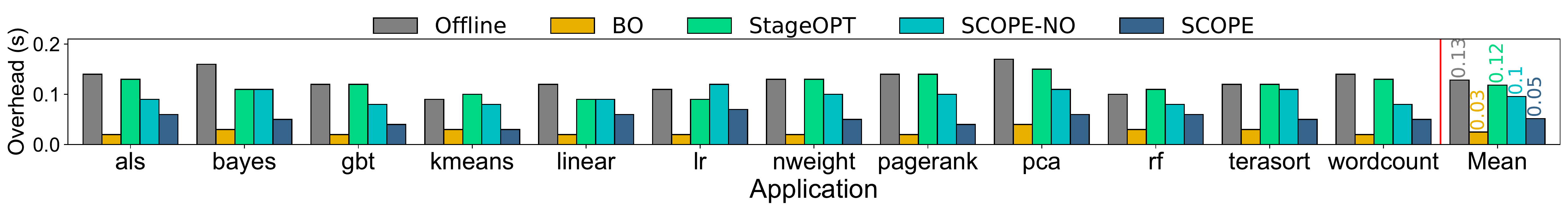} 
	\caption{Overhead of processing each sample for per reconfiguration averaged over all power constraints for each application. Lower is better.}\label{fig:overhead}
\end{figure*}

\subsection{RQ5: How sensitive is \SYSTEM{} to time intervals?}\label{sec:res-interval}

For all approaches that reconfigure dynamically, we evaluate how they
are affected by different time intervals (i.e., the time period
between reallocations). We sweep over different time intervals $[5,
10, 20, 30]$ seconds and choose the best time interval (i.e. the
interval that has the lowest violation rates across all power
constraints) for each method and each application.

Figure~\ref{fig:time_interval_sensitivity} summarizes the violation
rate over all power constraints and applications as a function of time
interval, where the x-axis is the different time intervals, and the
y-axis is the violation rate. 

Compared to \Offline{},
\BO{}, and \Stage{}, \SYSTEM{} has both the lowest violation rate and
the smallest variance at all intervals (\NO{} is the second best and
very close to \SYSTEM{}). It is because \SYSTEM{} and \NO{} construct
safe set and optimize latency together for each iteration, while \BO{}
optimizes for latency only, and \Stage{} separates expanding safe set
and optimizing latency into two separate stages. Critically, \Stage{}
only constructs the safe set at the beginning of execution, so if a
previously safe configuration becomes unsafe, then \Stage{} has no way to react.  
\SYSTEM{} is even better than \NO{} because it utilizes the locality
preserving operator to reconfigure more safely, which helps reduce the
overall violation rate. These results show that the novel safe
exploration framework of \SYSTEM{} is robust to the reconfiguring time
intervals.

\subsection{RQ6: How do different types of models perform?}\label{sec:res-surrogate}

\SYSTEM{} is a general framework compatible with any type of learning
model. Although Gaussian process regression is chosen as the learning
models in all our evaluations, we test the framework's generality by
using 4 common models: multi-layer
perceptron~\cite{ipek2006efficiently,ipek2005approach} (\MLP{}),
linear regression~\cite{ding2021generalizable} (\LIN{}), random
forest~\cite{nardi2019practical,chen2021efficient,roy2021bliss}
(\RF{}), and Gaussian process
regression~\cite{alipourfar2017cherrypick,patel2020clite} (\SYSTEM{}).
Note that for this experiment, we use the same type of models for
training both the safety and objective models.
Figure~\ref{fig:surrogates} summarizes the average violation rate and
speedup, where the x-axis is the application
and the y-axis is either the average violation rate or speedup, with
the last column Mean being the arithmetic mean over all applications.
We find that:
\begin{itemize}
\item \SYSTEM{} has the lowest violation rate compared to other
  learning models. On average, \SYSTEM{} has 4.54$\times$ lower
  violation rate than \MLP{}, 10.82$\times$ lower violation rate than
  \LIN{} and 12.0$\times$ lower violation rate than \RF{}. 
\item The speedup from using different learning models is relatively
  similar. \SYSTEM{} achieves 1.43$\times$ speedup over \MLP{}. \LIN{}
  and \RF{} are faster than \SYSTEM{} by 1\% and 7\% respectively.
  However, they are not good for being safe since both \LIN{} and
  \RF{} have significantly higher violations than \SYSTEM{}. 
\end{itemize}

\subsection{RQ7: What are the overheads?}\label{sec:res-overhead}

We report the overhead of processing each sample, which includes
updating the learning models to predict future system behavior and
choosing a new configuration. Figure~\ref{fig:overhead} shows the
average overhead of processing each sample for each application by
different approaches, and the last column Mean is the arithmetic mean
over all applications. \Static{}, \Oracle{} and \RAPL{} approaches
were omitted; \Static{} and \Oracle{} do not process samples or
reconfigure and \RAPL{} uses Intel's power control system to tune
parameters in the background and thus the overhead is not measurable.
In particular, we find that on average:
\begin{itemize}
\item \BO{} has the lowest overhead of 0.03s because it only trains
  one objective model without considering safety.
\item \SYSTEM{} has the second lowest average overhead and is better than
  \NO{} because of the locality preserving operator that reduces the
  number of configurations for inference in predicting safe set.
\item The overhead of \SYSTEM{}, which is 0.05s on average, is
  negligible given the total execution time. This is validated by 
  \SYSTEM{} achieving lowest latency even though we include overheads 
  in all experiments.  These overheads could be further
  reduced in future work by porting the resource manager to a
  lower-level language.
\end{itemize}

\section{Limitations}\label{sec:limitation}

We note the following limitations for this work:
\begin{itemize}
\item \SYSTEM{}, despite achieving the lowest violation rates and
  magnitudes, does not predict the complete safe set or provide
  any formal guarantees. Future work can explore
  formal guarantee for constructing the safe set.
\item Although Gaussian process regression provides confidence
  interval for prediction, such information was not used in \SYSTEM{}.
  Future work can explore techniques for incorporating
  uncertainty for safe exploration.
\item \SYSTEM{} currently uses fixed values for parameters such as
  time interval and the operator parameter $\gamma$ throughout
  execution. Future work can explore adaptively changing these
  parameters during execution to further reducing safety violations.
\end{itemize}

\section{Conclusion} \label{sec:conclusion}

This paper presents \SYSTEM{}, a resource manager that leverages a
novel safe exploration framework that dynamically allocate hardware 
resources in the execution space. \SYSTEM{} introduces a locality preserving
operator that reduces the violation rate and magnitudes
compared to prior work. 
We hope this work can inspire the computer systems community to build new
optimization tools that aid in exploration while preserving safety.

\bibliographystyle{unsrtnat}
\bibliography{refs}

\newpage

\section{Appendix}

This appendix contains the following items:
\begin{itemize}
    \item \cref{sec:app-percentiles} breaks down the results of violation rate, violation magnitude, and speedup in different safety constraint values. 
    \item \cref{sec:app-offline} shows how the number of samples used in training \Offline{} on its performance in optimizing for latency under power constraints.
\end{itemize}

\subsection{Evaluation on Different Safety Constraints} \label{sec:app-percentiles}
In~\cref{sec:res-power} and \cref{sec:res-latency}, we show the summarized results of each approach by taking the average across different power constraints. Here, we break down the results of each approach at each safety constraint to show how \SYSTEM{} is sensitive to the safety constraint.

\begin{figure*}[!htb]
	\centering
	\includegraphics[width=\linewidth]{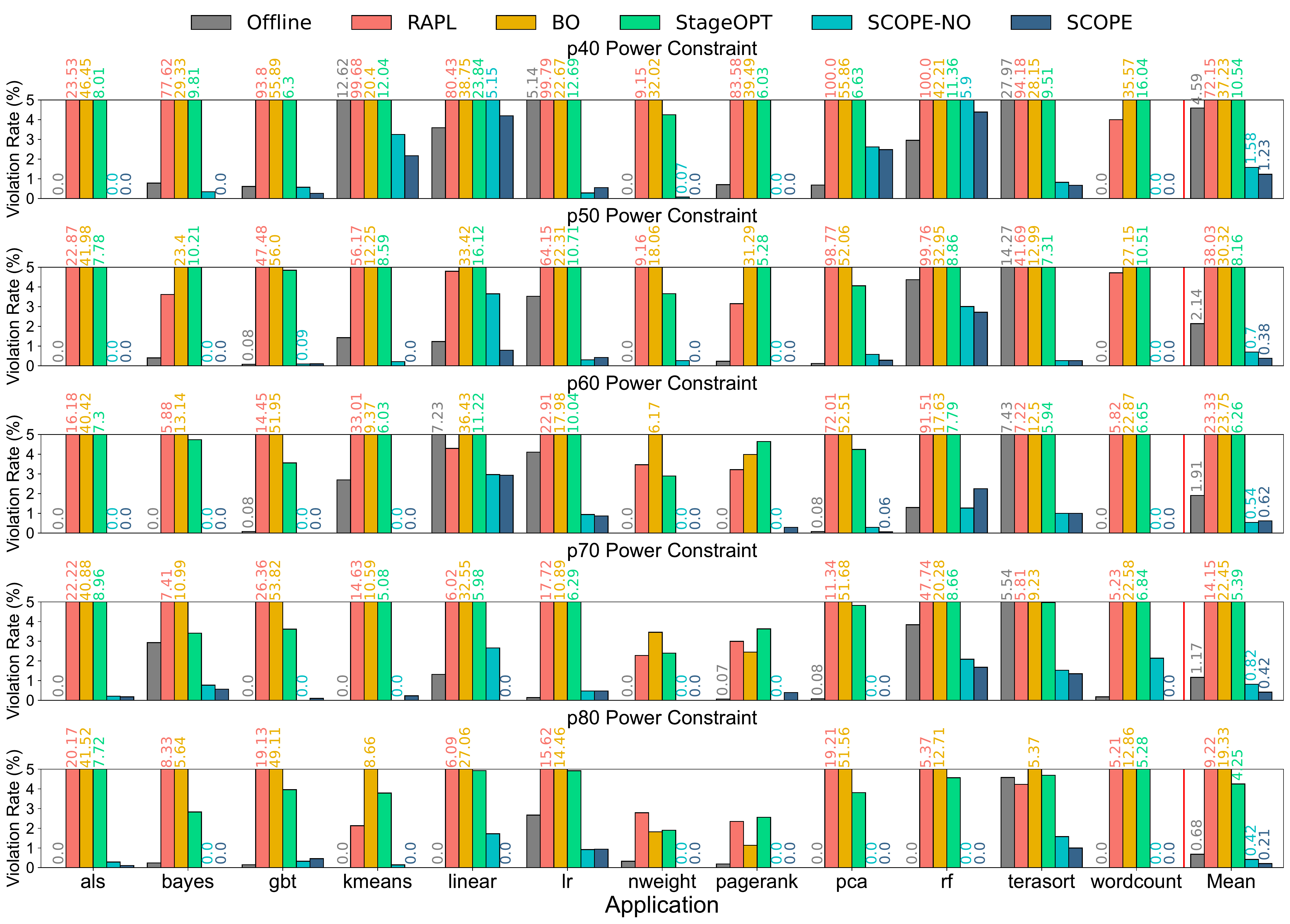} 
	\caption{Violation rates averaged over all random starting configurations for each power constraint and application. Lower is better.}\label{fig:percentile-violation-rate}
	\vspace{-0.1in}
\end{figure*}

\begin{figure*}[!htb]
	\centering
	\includegraphics[width=\linewidth]{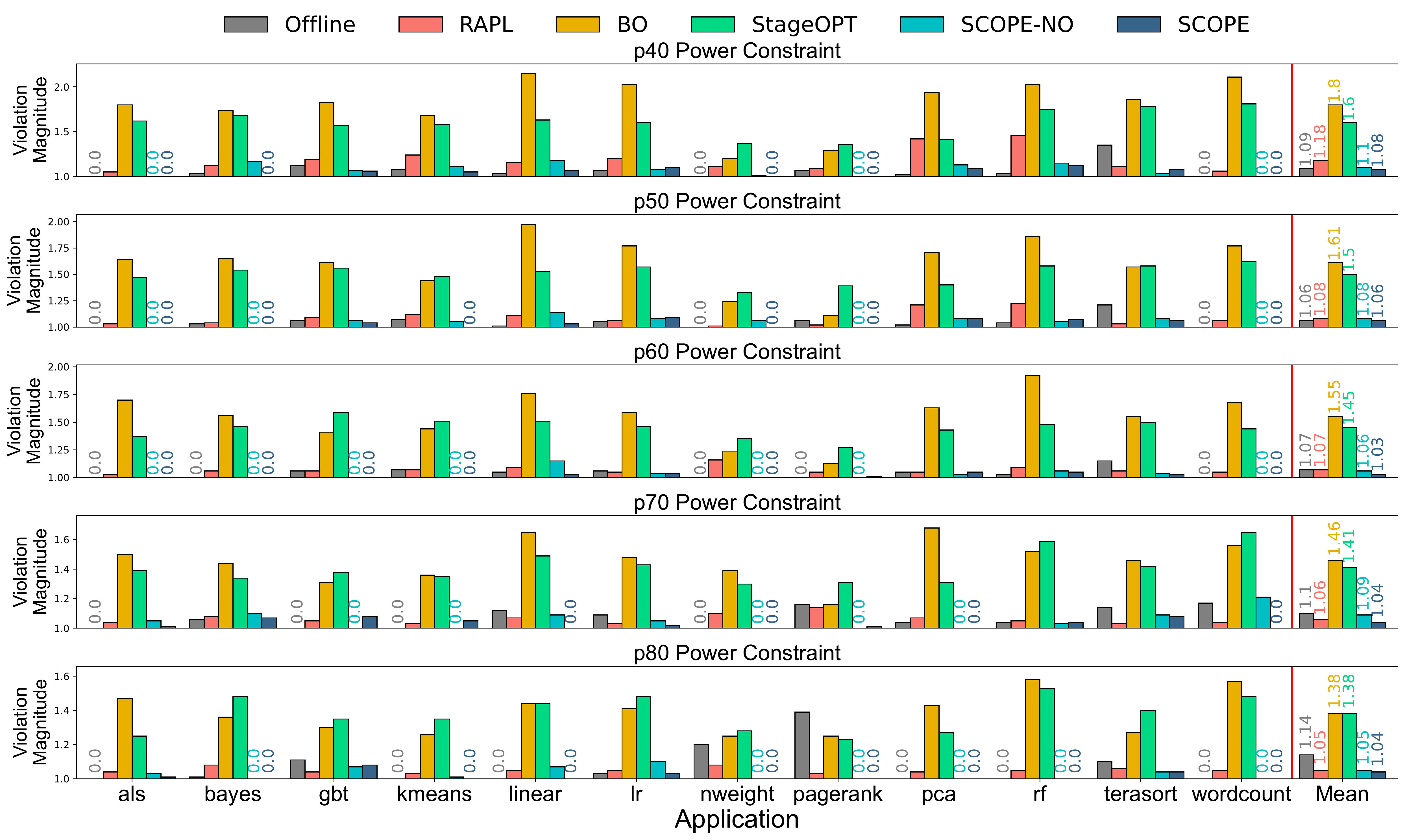} 
	\caption{Violation magnitudes averaged over all random starting configurations for each power constraint and application. Lower is better.}\label{fig:percentile-violation-magnitude}
	\vspace{-0.1in}
\end{figure*}

\paragraph{Violation rate and magnitude.} Figures~\ref{fig:percentile-violation-rate} and \ref{fig:percentile-violation-magnitude} show the violation rates and magnitudes for different approaches under different power constraints (p40-80), where the x-axis is the application, the y-axis is the violation rate or magnitude, the last column Mean is the arithmetic mean over all applications for Figures~\ref{fig:percentile-violation-rate} and arithmetic mean over all applications with non-zero violation magnitude for Figure~\ref{fig:percentile-violation-magnitude}, and we put the number on each bar to quantify the mean results. To visualize better, we cap the violation rate at 5\% in Figure~\ref{fig:percentile-violation-rate} and put numbers on the bars that are capped. 

We observe that \SYSTEM{} has the lowest violation rate and magnitude at almost all power constraints. These results explicitly show that \SYSTEM{} has a dominating advantage over other approaches no matter how we set the safety constraint.

\begin{figure*}[!htb]
	\centering
	\includegraphics[width=\linewidth]{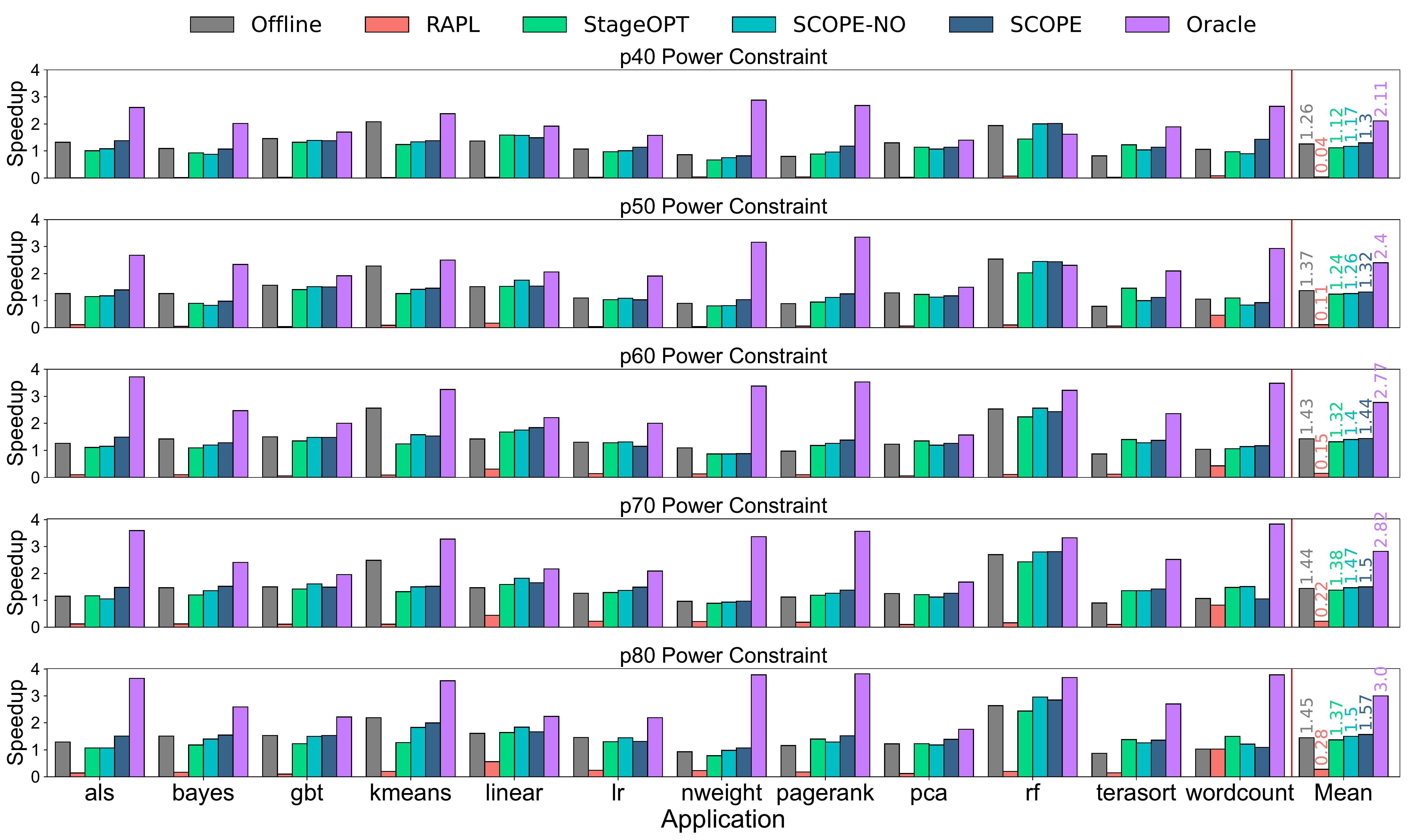} 
	\caption{Speedup averaged over all random starting configurations for each power constraint and application. Higher is better.}\label{fig:percentile-speedup}
	\vspace{-0.1in}
\end{figure*}

\paragraph{Speedup.} Figure~\ref{fig:percentile-speedup} shows the speedups over \Static{} for different approaches under different power constraints (p40-80), where the x-axis is the application, the y-axis is the speedup, the last column Mean is the arithmetic mean over all applications, and we put the number on each bar to quantify the mean results. We observe that \SYSTEM{} is second to \Oracle{} in all but the p50 power constraint, where \Offline{} has slightly higher speedup. This suggests that \SYSTEM{}'s dominating advantage is robust to the value of safety constraint.

\begin{figure*}[!htb]
	\centering
	\includegraphics[width=\linewidth]{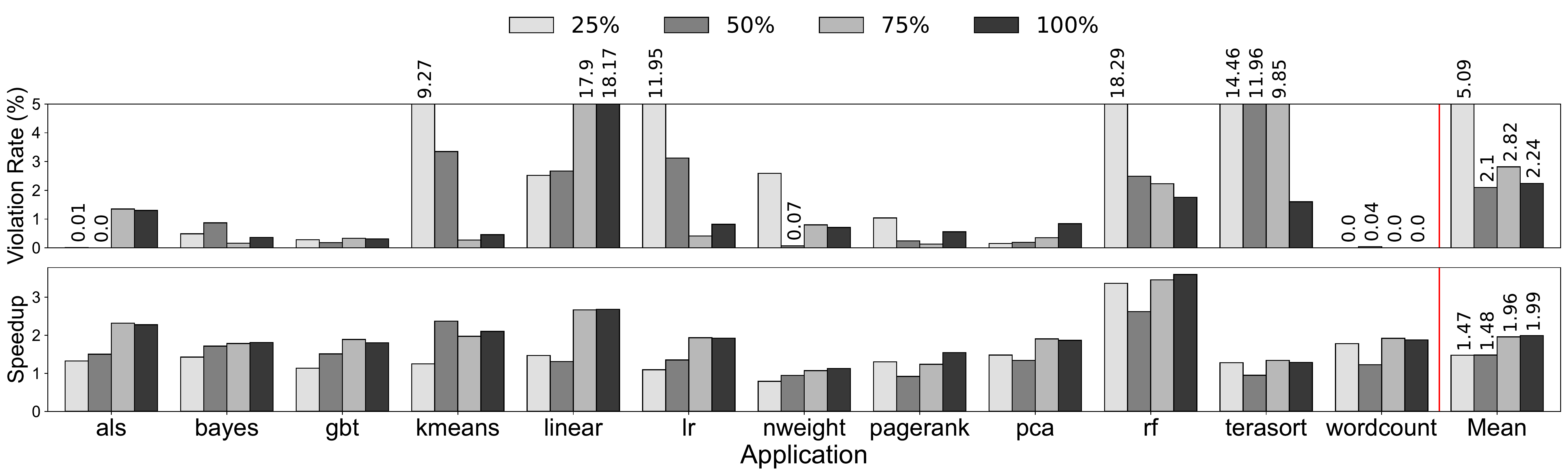} 
	\caption{Violation rates and speedup of \Offline{} using different amounts of samples, averaged over all power constraints for each application. Lower is better for violation rate. Higher is better for speedup.}\label{fig:offline-sweep}
	\vspace{-0.1in}
\end{figure*}

\subsection{Sensitivity Analysis for \Offline{} Approach} \label{sec:app-offline}
The \Offline{} approach that is compared against \SYSTEM{} in \cref{sec:comparison} samples 50\% of all possible hardware resource allocations from the target application and input for training. This section shows how \Offline{} is affected by the number of samples used. We sweep over different amount of samples, [25, 50, 75, 100]-\% of all possible hardware resource allocations using the target application and input. Figure~\ref{fig:offline-sweep} summarizes the average violation rate and speedup, where the x-axis is the application, the y-axis is the violation rate or speedup, the last column Mean is the arithmetic mean over all applications, and we put the number on each bar to quantify the mean results. We observe that using higher amount of samples has increased speedup for \Offline{}. However, using higher amount of samples does not necessarily lead to lower violation rate, further demonstrating the difficulty of generalization of the a priori collected samples.

\end{document}